\documentclass[12pt]{article}

\ifx\pdfoutput\undefined
\usepackage[dvips,bookmarks]{hyperref}
\else
\usepackage{hyperref}
\fi
\hypersetup{colorlinks=false,bookmarksopen,bookmarksnumbered,citecolor=blue,
   pdfstartview=FitH}

\usepackage[dvips]{graphicx}
\usepackage{latexsym}

\usepackage{amssymb,amsfonts,amsmath}
\usepackage{graphicx} 
\usepackage{indentfirst}

 \usepackage{bbm}

\parskip=\medskipamount

\newcommand {\cB}{{\cal B}}
\newcommand {\cC}{{\cal C}}

\newcommand {\cK}{{\cal K}}
\newcommand {\cL}{{\cal L}}
\newcommand {\cM}{{\cal M}}
\newcommand {\cN}{{\cal N}}

\newcommand {\cS}{{\cal S}}
\newcommand {\cT}{{\cal T}}
\newcommand {\cU}{{\cal U}}
\newcommand {\cV}{{\cal V}}

\newcommand{\bR}{{\bf R}}
\newcommand{\bS}{{\bf S}}
\newcommand{\bT}{{\bf T}}

\def\a{\alpha}

\def\b{\beta}

\def\d{\delta}

\def\g{\gamma}

\def\q{\theta}
\def\r{\rho}
\def\s{\sigma}

\def\F{\Phi}

\def\O{\Omega}

\def\S{\Sigma}

\def\rd{{\rm d}}
\def\ri{{\rm i}}
\def\re{{\rm e}}

\newcommand{\ad}{{\dot{\alpha}}}                           
\newcommand{\ve}{\varepsilon}                            
\newcommand{\DB}{\bar{D}}

\newcommand{\pa}{\partial}                           
\newcommand{\hf}{\frac12}

\newcommand{\be}{\begin{equation}}
\newcommand{\ee}{\end{equation}}
\newcommand{\bea}{\begin{eqnarray}}
\newcommand{\eea}{\end{eqnarray}}
\newcommand{\non}{\nonumber}
\newcommand{\1}{{\underline{1}}}
\newcommand{\2}{{\underline{2}}}

\def\dt#1{{\buildrel {\hbox{\LARGE .}} \over {#1}}}    

\newcommand{\bm}[1]{\mbox{\boldmath$#1$}}

\def\double #1{#1{\hbox{\kern-2pt $#1$}}}

\newcommand{\alu}{{\underline{\a}}}
\newcommand{\beu}{{\underline{\b}}}
\newcommand{\gau}{{\underline{\g}}}
\newcommand{\deu}{{\underline{\d}}}

\newcommand{\qb}{{\bar{\theta}}}

\newcommand{\bmL}{{\bm L}}
\newcommand{\bmR}{{\bm R}}
\newcommand{\bpl}{{\boxplus}}
\newcommand{\bmn}{{\boxminus}}
\newcommand{\opl}{{\oplus}}
\newcommand{\omn}{{\ominus}}

\newif\ifdtup

\def\de{{\nabla}}                                         
\def\cm{{\cal M}}

\def\deb{{\bar{\de}}}

\def\pp{+\hspace{-0.04in}+}

\newcommand{\bsubeq}{\begin{subequations}}
\newcommand{\esubeq}{\end{subequations}}

\begin{document}

\begin{titlepage}
\begin{flushright}
UMD-PP-09-046\\
November, 2009\\
\end{flushright}
\vspace{5mm}

\begin{center}
{\Large \bf 2D ${\cal{N}}=(4,4)$
superspace supergravity and bi-projective superfields}\\ 
\end{center}

\begin{center}

{\bf
Gabriele Tartaglino-Mazzucchelli\footnote{gtm@umd.edu}
} \\
\vspace{5mm}

{\it Center for String and Particle Theory,
Department of Physics, 
University of Maryland\\
College Park, MD 20742-4111, USA}
~\\

\end{center}
\vspace{5mm}

\begin{abstract}
\baselineskip=14pt

We propose a new superspace formulation for ${\cal{N}}=(4,4)$ conformal supergravity in 
two dimensions.
This is based on a  geometry
where the structure group of the curved superspace
is chosen to be SO(1,1)$\times$SU(2)$_L\times$SU(2)$_R$.
The off-shell supergravity multiplet possesses 
super-Weyl transformations generated by 
an unconstrained real scalar superfield.
The new supergravity formulation turns out to be an extension of the minimal multiplet
introduced  in 1988 by Gates {\it et.~al}.
and it allows the existence of various off-shell matter supermultiplets.
Covariant twisted-II and twisted-I multiplets respectively describe  
the  field strength of an Abelian vector multiplet 
and its prepotential.
Moreover, we introduce covariant bi-projective superfields.
These define a large class of matter multiplets coupled to 2D ${\cal{N}}=(4,4)$ 
conformal supergravity.
They are the analogue of the covariant projective superfields recently  introduced for 4D and 5D 
matter-coupled supergravity but they differ by the fact that bi-projective superfields 
are defined with the use of two $\mathbb{C}P^1$ instead of one.
We conclude by giving a manifestly locally supersymmetric and super-Weyl invariant 
action principle in bi-projective superspace.

\end{abstract}
\vspace{1cm}

\vfill
\end{titlepage}

\newpage
\renewcommand{\thefootnote}{\arabic{footnote}}
\setcounter{footnote}{0}


\tableofcontents{}
\vspace{1cm}
\bigskip\hrule


\section{Introduction}
\setcounter{equation}{0}

In revealing the off-shell structures of supersymmetric field theories a most natural framework is 
provided by superspace.
This can offer a formalism to build
general supersymmetric models with covariance fully guaranteed
that is especially important, for example, in studying supergravity.
An adequate superspace formalism is also 
powerful in the analysis of the quantum behavior of globally and locally supersymmetric 
field theories and it proves to be a unique
tool in understanding the target space geometry of supersymmetric non-linear sigma-models.
These statements have probably their best and simplest explanations
 in the case of 4D $\cN=1$ supersymmetry (see \cite{SUPERSPACE,BK,WB} for reviews).
In the case of extended supersymmetry off-shell formulations using superspace, when 
possible, are less simple and complete prescriptions are, in our opinion, still to be found.

Exemplary is the case of supersymmetry with eight real supercharges 
in its most studied form: 4D $\cN=2$ supersymmetry.
In this case, the study of supersymmetric multiplets and invariants
naturally leads to the introduction of extended superspace 
coordinates related to the SU(2) automorphism  group
of the supersymmetry algebra \cite{Rosly,harm1,KLR}. 
Then, invariant sub-superspaces emerge
and one can  treat general multiplets including off-shell charged hypermultiplets.
These, to close off-shell  supersymmetry without central charges \cite{hyper,RocekSiegel}, 
have an infinite number of auxiliary fields \cite{harm1,harm,KLR,LR,Rey}.

In the literature, two superspace formalism have been introduced to study 
4D $\cN=2$ supersymmetric field theories. 
They go under the names of harmonic superspace (HS) \cite{harm1,harm}
and projective superspace (PS) \cite{KLR,LR}.
The methods make use of the two equivalent superspaces
${\mathbb R}^{4|8}\times S^2$ and ${\mathbb R}^{4|8}\times {\mathbb C}P^1$ respectively,
however, they differ in the structure of the off-shell supermultiplets used 
and the supersymmetric action principle chosen. 
Due to their differences, the two approaches often prove to be  complementary 
to each other\footnote{For global supersymmetry, the relationship 
between the harmonic and projective superspace has been described
in \cite{K-double}. See also \cite{JainSiegel} for a recent discussion.}.
In this paper we will focus on projective superspace.

Projective superspace was first introduced to study  globally supersymmetric non-linear 
sigma-models providing, since then, a powerful generating formalism to build new hyper-K\"ahler 
metrics\footnote{For a review on this subject, and a partial list of references,
see \cite{LRtwistor} where a nice discussion of the relationship between twistor spaces and 
projective superspace is given.}
 \cite{KLR,GHR,HKLR,LR}.
PS has proved to be a useful approach in studying supersymmetric field theories also
in 5D \cite{KL} and 6D \cite{GL,GPT}. 
 Superconformal field theories in PS have been 
 described by Kuzenko in four and five 
 dimensions \cite{K,K2} providing a starting point for curved 
 extensions.\footnote{Building on the superconformal projective
multiplets of \cite{K,K2},  for a curved geometry 
projective superfields 
were first introduced
in studying field theory
in 5D $\cN=1$ anti-de Sitter superspace \cite{KT-M}.}
In the supergravity case, we recently proposed a PS formalism first in five
\cite{KT-M_5D,KT-M_5Dconf} and then in four dimensions \cite{KLRT-M_4D-1,KLRT-M_4D-2}
(see \cite{KT-M_5DconfFlat,KT-M_4DconfFlat,Kdual,KT-M-normal} for recent developments and 
applications).
This PS approach
is conceptually based on two main ingredients: 
(i) a constrained superspace geometric formulation 
of the Weyl multiplet of conformal supergravity, which is based on ``standard'' 
Wess-Zumino superspace \cite{WZ-s} 
techniques;\footnote{For the Weyl multiplet of 5D $\cN=1$ conformal 
supergravity  \cite{5DN1-Weyl} we gave a superspace formulation in \cite{KT-M_5Dconf}.
In the 4D $\cN=2$ conformal case we made use in \cite{KLRT-M_4D-1} of the Grimm's 
superspace geometry \cite{Grimm} while in \cite{KLRT-M_4D-1} we considered the
Howe's formulation \cite{Howe}; the supergravity of \cite{Grimm} is obtained by a partial 
gauge--fixing of the geometry of \cite{Howe} but they both describe the Weyl multiplet 
\cite{4DN2-Weyl}.}
(ii) the existence of 
covariant projective multiplets
 and a locally supersymmetric and super-Weyl invariant 
action principle in PS that are consistently defined on the curved
geometry of point (i).
These ingredients allow a covariant {\emph{off-shell}} setting 
for general 5D $\cN=1$ and 4D $\cN=2$ supergravity-matter systems 
similar to the  Wess-Zumino superspace approach to 4D $\cN = 1$ supergravity.
It is worth mentioning that  a prepotential formulation for 4D $\cN=2$ conformal supergravity
was given in harmonic superspace twenty years ago \cite{Galperin:1987em}. 
However, its  relationship to the standard, curved superspace geometrical methods has not yet
been elaborated in detail.
A synthesis of HS and PS, could possibly provide a coherent superspace description 
of 4D $\cN = 2$ supergravity, similar to the famous Gates-Siegel 
prepotential approach to the 4D $\cN = 1$ case \cite{SG}.

Projective superspace has been introduced also for two-dimensional 
$\cN=(4,4)$ supersymmetry \cite{GHR,BusLinRoc,RSS,LR-biProj}.
The 2D case is interesting and peculiar for many reasons.
First of all, once the number of supercharges is 
fixed, decreasing the spece-time dimensions the 
number of inequivalent multiplets can increase due to ``twisting'' phenomena.
In this regard, the 2D case is exemplar
(see \cite{GatesKetov} for a discussion of the 2D $\cN=(4,4)$ case).
For instance, in 2D the PS approach 
makes the explicit use of two ${\mathbb{C}}P^1$ coordinates
leading to a richer class of multiplets and sigma-models than in 4D.
Moreover, 2D supersymmetry clearly has an important role in the 
classification of both superconformal field theories \cite{ConformalBOOKS} 
and string theory \cite{StringBOOKS}.
Based on the seminal paper \cite{GHR}, the recent observation that 
generalized complex geometry (GCG)
 \cite{GCG} arises as the target space geometry 
of 2D supersymmetric non-linear sigma-models, has also renewed the interest on
2D superspace techniques especially for the $\cN=(2,2)$ cases
\cite{GHR,22superspace,GCGsuperspace-1,GCGsuperspace-2}.
A main interest in GCG is due to its importance 
in string theory compactification with fluxes
 (see \cite{Grana} for a  review).
The 2D $\cN=(4,4)$ case has been less explored but it could be 
interesting, for example, to generate new bi-Hermitian and generalized hyper-K\"ahler geometries 
(besides the works \cite{GHR,BusLinRoc,RSS,LR-biProj}
see \cite{GotLin} for a recent analysis in 2D $\cN=(2,2)$ superspace).

With the previous observations in mind, this work is focused 
on the development of new superspace techniques for 2D $\cN = (4, 4)$ supergravity.
In particular there are two main goals of the paper:

(a) a new superspace formulation for 2D $\cN= (4,4)$ conformal 
supergravity, and;

(b) a two-dimensional generalization of the four and five-dimensional curved projective 
superspace approach of \cite{KT-M_5D,KT-M_5Dconf,KLRT-M_4D-1,KLRT-M_4D-2}.

In building up the 2D $\cN=(4,4)$ curved PS techniques  
we follow the same principles used in 4D and 5D: 
(i) we identify a Wess-Zumino superspace  constrained geometric formulation 
of 2D conformal supergravity;
(ii) we introduce covariant supermultiplets
 and a locally supersymmetric and super-Weyl invariant 
action principle in 2D PS.
Generalizing the flat case of \cite{BusLinRoc,RSS,LR-biProj},
2D curved PS depends on extra ${\mathbb{C}}P^1\times{\mathbb{C}}P^1$ coordinates;
for this reason we call the superspace and supermultiplets 
\emph{bi-projective}.\footnote{For 2D $\cN=(4,4)$ supersymmetry
HS has been introduced in \cite{IvanovSutulin}.
In analogy to bi-projective superspace \cite{BusLinRoc},
 the 2D HS makes use of two sets of 
harmonics.
A prepotential formulation for 2D $\cN=(4,4)$ conformal supergravity has been given
in bi-harmonic superspace \cite{BellucciIvanov}.  
A detailed analysis of 
the relationship between 2D PS and HS superspaces,
both in the flat and curved cases, would be useful.}

Twenty years ago Gates {\emph{et.~al.}} presented a superspace formulation for 
2D $\cN=(4,4)$ minimal supergravity \cite{2DN4SG} (see \cite{2DSGearly} for earlier  
components results). 
This was based on the gauging of a SO(1,1)$\times$SU(2)$_\cV$ tangent space
group and a suitable irreducible set of constraints on the torsion of the
curved superspace.
Another important feature 
was the realization 
of the superconformal group.
The super-Weyl transformations, that preserve the minimal torsion constraints of \cite{2DN4SG}, 
are generated by
two real superfields $\bS,\,\bS_{ij}=\bS_{ji}$ through the following infinitesimal 
variation of the spinor covariant derivative\footnote{Here $\de_{\a i}$ are the complex spinor 
covariant derivatives, $\cm$ and $\cV_{kl}$ are respectively 
the Lorentz and SU(2)$_\cV$ generators.
The $\bS,\bS_{ij}$ superfields are not independent 
satisfying the differential constraint
$\de_{\a i}\bS_{kl}=-\hf(\g^3)_\a{}^\b C_{i(k}\de_{\b k)}\bS$.
More discussions
are in the body of the paper.}
\bea
\tilde{\d} \de_{\a i}&=&
\hf\bS\de_{\a i}
+(\g^3)_\a{}^\b\bS_i{}^{j}\de_{\b j}
 +(\g^3)_\a{}^\g(\de_{\g i}\bS)\cM
 +(\de_{\a}^{k}\bS)\cV_{ik}
~.
\label{SW-000}
\eea
From the previous equation, it turns out that the lowest component of $\bS$, 
$\bS|:=\bS(x,\theta)|_{\theta=0}$, generates
local scale transformations. The second term in (\ref{SW-000}) induces a chiral 
SU(2)$_\cC$ transformation 
generated by 
$\bS_{ij}|$.
Special supersymmetry transformations are generated by $\de_{\a i}\bS|$ 
and so on \cite{2DN4SG}.
Moreover, looking at (\ref{SW-000}) it is clear that the $E_{\a i}{}^A$ supervielbein 
does not transform 
homogeneously under super-Weyl transformations.
A natural question which arise from the previous observations is: 
does a 2D $\cN=(4,4)$ geometry with ``homogeneous''
super-Weyl transformations exists?
The answer is an easy yes. We find one by enlarging the minimal tangent space group 
with the inclusion of SU(2)$_\cC$ transformations.\footnote{For reasons that will become clear in 
section 2, we will refer to the structure group of the new extended supergravity
formulation as SO(1,1)$\times$SU(2)$_L\times$SU(2)$_R$.}
Then, it is easy to find a set of torsion constraints for the curved superspace 
that are preserved by super-Weyl transformations 
generated by a single real unconstrained scalar superfield.
Of course the new 
supergravity
is reducible and, upon gauge fixing,
gives the minimal multiplet 
of \cite{2DN4SG}.

The important point is that 
by using the extended supergravity formulation one can easily couple the 
geometry to many matter multiplets.
Here we present covariant 
 twisted-I (TM-I) \cite{Gnew2D,GHR} 
and twisted-II (TM-II) \cite{TM-II-IK} matter multiplets.
Moreover, covariant bi-projective superfields are consistently defined;
this is one of the main results of the paper.
An advantage of the new supergravity is that 
all the matter multiplets considered possess homogeneous  super-Weyl transformations
in the extended geometry.
In the minimal case, the matter multiplet's super-Weyl transformations,
that we will prove to be in-homogeneous and thus a bit more ``tricky'', 
are easily spelled out by the details
of the reduction from the extended geometry to the minimal one.
The results contained in this paper then
explain and extend the analysis of \cite{2DN4SG} and, more importantly,
 give a covariant prescription to study general 2D $\cN=(4,4)$ conformal 
 supergravity-matter systems.

This paper is organized as follows.
In section 2 we describe the geometry of the new extended 
SO(1,1)$\times$SU(2)$_L\times$SU(2)$_R$ curved superspace.
 We include the finite super-Weyl transformations 
and comment about the gauge fixing to the minimal supergravity multiplet.
Section 3 contains the coupling to
a vector multiplet.
We observe that the geometry allows a coupling to an irreducible multiplet which has field 
strengths describing a covariant TM-II multiplet. 
We then describe a useful solution of the covariant TM-II
constraints and observe how a covariant TM-I matter multiplet emerges as a prepotential for the 
TM-II. We then discuss again on the gauge fixing to minimal supergravity.
Section 4 is devoted to the definition of 2D curved bi-projective superspace.
We define a large class of covariant bi-projective superfields and formulate
a locally supersymmetric and super-Weyl invariant action principle.
Section 5 contains some concluding observations.
This paper is accompanied by three technical appendices. 
In appendix A we collect our 2D conventions.
Appendix B summarizes the solution of the Bianchi identities 
for the supergravity geometry of subsection \ref{ExtendedSUGRA}.
Appendix C contains a derivation of eq. (\ref{dedeU}) which is crucial for the analysis 
of section 4.


\section{2D $\cN=(4,4)$ conformal supergravity geometries}
\setcounter{equation}{0}
\label{SUGRA}

In this section we present a new covariant superspace description of 2D $\cN=(4,4)$ conformal 
supergravity based on an extension of the minimal multiplet of 
Gates {\it et.~al.} \cite{2DN4SG}.
There are two main differences between the two formulations. 
The first is the choice of the supergravity structure group. In the minimal case this was 
SO(1,1)$\times$SU(2)$_\cV$, in the present case we make use of
SO(1,1)$\times$SU(2)$_L\times$SU(2)$_R$.
The second difference 
are the super-Weyl transformations.
In the minimal case these 
 are generated by a twisted-II multiplet 
\cite{2DN4SG} while in the present, extended formulation  a real unconstrained scalar superfield
 is the transformation parameter.
As we will see, the minimal multiplet can be obtained by a partial gauge fixing of the super-Weyl 
and SU(2) transformations.

\subsection{New SO(1,1)$\times$SU(2)$_L\times$SU(2)$_R$ superspace geometry}
\label{ExtendedSUGRA}

Consider a curved 2D $\cN=(4,4)$ superspace, which we will denote by $\cM^{2|4,4}$.
This is locally parametrized by coordinates
$z^M=(x^m,\q^{\mu \imath},\qb^\mu_\imath)$ where $m=0,1$, $\mu=+,-$ and $\imath=\1,\2$.
In the light-cone coordinates the superspace is locally parametrized by
$z^M=(x^{\pp},x^{=},\q^{+\imath},\qb^+_\imath,\q^{-\imath},\qb^-_\imath)$ where 
$x^{\pp}=\hf(x^1+x^0)$ and $x^{=}=\hf(x^1-x^0)$. 
The Grassmann variables are related one to each other by the complex conjugation rule
$(\q^{\mu \imath})^*=\qb^\mu_{\imath}$ (see appendix A for our 2D conventions).
We choose the supergravity structure group to be SO(1,1)$\times$SU(2)$_L\times$SU(2)$_R$
and let $\cm$, $\bmL_{ij}$, $\bmR_{ij}$ be the corresponding Lorentz, SU(2)$_L$
and SU(2)$_R$ generators.
The label $L$ or $R$  are associated to a SU(2) group that respectively acts non-trivially 
only on the left or right light-cone sectors.

The covariant derivatives 
$\de_{{A}} =(\de_{{a}}, \de_{\a i},\deb_\a^i)$ 
(or $\de_{{A}} =(\de_{\pp},\de_{=}, \de_{+ i},\deb_+^i,\de_{- i},\deb_-^i)$)
are
\begin{subequations}
\bea
\de_{A}&=&E_{A}
~+~\O_{A}\,\cm
~+~(\Phi_L)_{A}{}^{kl}\,\bmL_{kl}
~+~(\Phi_R){}_{A}{}^{kl}\,\bmR_{kl}
~,\label{CovDev-3}
\\
\de_{A}&=&E_{A}
~+~\O_{A}\,\cm
~+~(\Phi_\cV)_{A}{}^{kl}\,\cV_{kl}
~+~(\Phi_\cC){}_{A}{}^{kl}\,\cC_{kl}
~.\label{CovDev-4}
\eea
\end{subequations}
Here $E_{{A}}= E_{{A}}{}^{{M}}(z) \pa_{{M}}$ is the supervielbein, 
with $\pa_{{M}}= \pa/ \pa z^{{M}}$,
$\O_{{A}}(z)$ is the Lorentz connection, 
 $(\Phi_L)_{{A}}{}^{kl}(z)$ and $(\Phi_R)_{{A}}{}^{kl}(z)$ are  
 the ${\rm SU}(2)_L$ and ${\rm SU}(2)_R$ connections, respectively.
We have also introduced the generators $\cV_{kl}$ and $\cC_{kl}$ defined by
\bea
\cV_{kl}=\bmL_{kl}+\bmR_{kl}~,~~
\cC_{kl}=\bmL_{kl}-\bmR_{kl}~,~~~
\bmL_{kl}=\hf(\cV_{kl}+\cC_{kl})~,~~
\bmR_{kl}=\hf(\cV_{kl}-\cC_{kl})~,
\eea
with $(\Phi_\cV)_{{A}}{}^{kl}(z)$ and $(\Phi_\cC)_{{A}}{}^{kl}(z)$ their connections.

The action of the Lorentz generator on the covariant derivatives is as follow
\begin{subequations}
\bea
&&{[} {\cal M} , \nabla_{\a  i} {]} = \hf 
(\g^3)_{\a} {}^{\b} \nabla_{\b  i}    ~,~~~
{[} {\cal M} , \deb_{\a }^i  {]} = \hf 
(\g^3)_{\a} {}^{\b} \deb_{\b}^i~,~~~
{ [}  {\cal M} , \nabla_{a}  {]} =
\ve_{ab}\nabla^b~,     
\label{Lorentz1}
\\
&&{[} {\cal M} , \nabla_{\pm  i} {]} = \pm\hf  \nabla_{\pm  i}    ~,~~~
{[} {\cal M} , \deb_{\pm }^i  {]} = \pm\hf \deb_{\pm}^i~,~~~
{ [}  {\cal M} , \nabla_{\buildrel \pp \over =}{]}
 =\pm\nabla_{\buildrel\pp \over =}
~.     
\label{Lorentz2}
\eea
\end{subequations}
The action of the SU(2)$_L$ and SU(2)$_R$  generators on the covariant 
derivatives is
\begin{subequations}
\bea
{[}  {\bmL}{}_{kl} , \nabla_{+ i} {]} = \hf C_{i(k} \nabla_{+ l)} ~,~~
{[}  {\bmL}{}_{kl} , \deb_{+}^i  {]} = -\hf \d^{i}_{(k} \deb_{+ l)} ~,~&&
{[}  {\bmL}{}_{kl} , \nabla_{- i} {]} = {[}  {\bmL}{}_{kl} , \deb_{-}^{ i} {]}=0 ~,~~~~~~
\label{L_kl}
\\
{[}  {\bmR}{}_{kl} , \nabla_{- i} {]} = \hf C_{i(k} \nabla_{- l)} ~,~~
{[}  {\bmR}{}_{kl} , \deb_{-}^i  {]} = -\hf \d^{i}_{(k} \deb_{- l)} ~,~&&
{[}  {\bmR}{}_{kl} , \nabla_{+ i} {]} = {[}  {\bmR}{}_{kl} , \deb_{+}^{ i} {]}=0 ~,~~~~~~
\label{R_kl}
\eea
\end{subequations}
and $\cV_{kl}$ and $\cC_{kl}$ satisfy
\begin{subequations}
\bea
&&{[}  {\cV}{}_{kl} , \nabla_{\a i} {]} = \hf C_{i(k} \nabla_{\a l)} 
~,~~~
{[}  {\cV}{}_{kl} , \deb_{\a}^i  {]} = -\hf \d^{i}_{(k} \deb_{\a l)} 
~,
\label{V_kl}
\\
&&{ [} {\cC}_{kl} , \nabla_{\a i}  {]} = \hf (\g^3)_\a{}^\b C_{i(k} \nabla_{\b l)} 
~,~~~
{[} {\cC}_{kl} , \deb_{\a}^i {]} = -\hf (\g^3)_\a{}^\b \d^{i}_{(k} \deb_{\b l)} 
~.
\label{C_kl}
\eea
\end{subequations}
Moreover, it holds 
$[\bmL_{kl},\de_a]=[\bmR_{kl},\de_a]=[\cV_{kl},\de_a]=[\cC_{kl},\de_a]=0$.
From the previous equations it is clear that the operator $\cV_{kl}$ generates a diagonal 
SU(2)$_\cV$ 
subgroup inside SU(2)$_L\times$SU(2)$_R$ and $\cC_{kl}$ generates chiral SU(2)$_\cC$ 
transformations.
The algebra of commutators of the structure group generators is given in appendix A eq.
(\ref{[MMLR]})--(\ref{[CC]}).

The supergravity gauge group is given by local general coordinate and tangent space 
transformations of the form 
\begin{subequations}
\bea
&&\d_\cK \de_{{A}} =[\cK  , \de_{A}]~,
\label{SUGRAgauge1}
\\
&&\cK = K^{{C}} \de_{{C}} +K\cm
+(K_L)^{kl} \bmL_{kl} 
+(K_R)^{kl} \bmR_{kl} ~,
\label{SUGRAgauge2}
\\
&&\cK = K^{{C}} \de_{{C}} +K\cm
+(K_\cV)^{kl} \cV_{kl} 
+(K_\cC)^{kl} \cC_{kl} ~,
\label{SUGRAgauge3}
\eea
\end{subequations}
with the gauge parameters
obeying natural reality conditions, but otherwise  arbitrary superfields. 
Given a tensor superfield $\cU(z)$, with its indices suppressed, 
it transforms as:
\bea
\d_\cK \cU = \cK\, \cU~.
\label{tensor-K}
\eea

The covariant derivatives algebra has the form
\begin{subequations}
\bea
{[}\de_{{A}},\de_{{B}}\}&=&T_{{A}{B}}{}^{{C}}\de_{{C}}
+R_{{A}{B}}\cm
+(R_L)_{{A}{B}}{}^{kl}\bmL_{kl}
+(R_R){}_{{A}{B}}{}^{kl}\bmR_{kl}
~,
\label{algebra-4-0}
\\
{[}\de_{{A}},\de_{{B}}\}&=&T_{{A}{B}}{}^{{C}}\de_{{C}}
+R_{{A}{B}}\cm
+(R_\cV)_{{A}{B}}{}^{kl}\cV_{kl}
+(R_\cC){}_{{A}{B}}{}^{kl}\cC_{kl}
~,
\label{algebra-4}
\eea
\end{subequations}
where  $T_{AB}{}^C$ is the torsion, $R_{AB}$ is the Lorentz curvature,
 $(R_L)_{AB}{}^{kl},\,(R_R)_{AB}{}^{kl}$ are the SU(2)$_L\times$SU(2)$_R$ curvatures
 that have been recombined in the second line as 
  $(R_\cV)_{AB}{}^{kl},\,(R_\cC)_{AB}{}^{kl}$ 
 in terms of the generators $\cV_{kl},\cC_{kl}$.

For the remainder of this section we will always use the $\cV_{kl},\,\cC_{kl}$ parametrization of the  
SU(2)$_L\times$SU(2)$_R$ group. 
In that basis it will be trivial to see the reduction of our supergravity multiplet to the minimal one 
of \cite{2DN4SG} where the structure group of the curved superspace was chosen to be 
SO(1,1)$\times$SU(2)$_\cV$.

We impose the following constraints on the torsion
\begin{subequations}
\bea
&T_{\a i}{}_\b^j{}^c=2\ri\d_i^j(\g^{c})_{\a\b}~,~~~
T_{\a i}{}_{\b j}{}^{c}=0~,
~~~~~~
&{\rm (dimension~0)}
\label{constr-0}
\\
&T_{\a i}{}_{\b j}{}^{\g k}
=T_{\a i}{}_{\b j}{}^{\g}_k
=T_{\a i}{}_{b}{}^c=0~,
~~~~~~&{\rm (dimension~1/2)}
\label{constr-1/2}
\\
&\d^\b_\g T_{a \b (j}{}^\g_{k)}=(\g^3)_\g{}^\b T_{a \b (j}{}^\g_{k)}=T_{ab}{}^{c}=0~,
~~~~~~
&{\rm (dimension~1)}
\label{constr-1}
\eea
\end{subequations}
along with their complex conjugates.

The solution of the Bianchi identities based on the constraints (\ref{constr-0})--(\ref{constr-1})
is given for the interested reader in appendix \ref{SUGRA-Bianchi}.
Here we collect the main results.

The algebra of covariant derivatives based on (\ref{constr-0})--(\ref{constr-1}) results to be
\begin{subequations}
\bea
\{\de_{\a i},\de_{\b j}\}&=&
-4\ri\Big(C_{ij}C_{\a\b}N
-(\g^3)_{\a\b}Y_{ij}
\Big)\cm
+4\ri \Big((\g^3)_{\a\b}N-(\g^a)_{\a\b}A_a\Big)\cV_{ij}
\non\\
&&
+2\ri C_{ij}C_{\a\b}Y^{kl}\cV_{kl}
+2\ri (\g^3)_{\a\b}Y_{(i}{}^{k}\cC_{j)k}
-4\ri (\g_a)_{\a\b}\ve^{ab}A_b\cC_{ij}
~,
\label{Algebra-1.1}
\\
\{\deb_\a^i,\deb_\b^j\}&=&
4\ri\Big(C^{ij}C_{\a\b}\bar{N}
-(\g^3)_{\a\b}\bar{Y}^{ij}
\Big)\cM
-4\ri \Big((\g^3)_{\a\b}\bar{N}+(\g^a)_{\a\b}\bar{A}_a\Big)\cV^{ij}
\non\\
&&
-2\ri C^{ij}C_{\a\b}\bar{Y}_{kl}\cV^{kl}
+2\ri (\g^3)_{\a\b}\bar{Y}^{(i}{}_{k}\cC^{j)k}
-4\ri (\g_a)_{\a\b}\ve^{ab}\bar{A}_b\cC^{ij}
~,
\label{Algebra-1.1c}
\\
\{\de_{\a i},\deb_\b^j\}&=&
2\ri\d_i^j(\g^a)_{\a\b}\de_a
-4\ri\Big(C_{\a\b}\big(\d_i^j\cT+\ri\cT_{i}{}^{j}\big)
-(\g^3)_{\a\b}\big(\ri \d_i^j\cS+\cS_{i}{}^{j}\big)
\Big)\cm
\non\\
&&
+4\Big(
\ri(\g^3)_{\a\b}\cT
+C_{\a\b}\cS
+(\g_a)_{\a\b}\cB^a
\Big)\cV_{i}{}^j
+2\d_i^j\Big(
(\g^3)_{\a\b}\cT^{kl}
+\ri C_{\a\b}\cS^{kl}
\Big)\cV_{kl}
\non\\
&&
+2C_{\a\b}\cT_i{}^k\cC^j{}_{k}
+2C_{\a\b}\cT^{jk}\cC_{ik}
+2\ri (\g^3)_{\a\b}\cS_i{}^k\cC^j{}_{k}
+2\ri (\g^3)_{\a\b}\cS^{jk}\cC_{ik}
\non\\
&&
+4 (\g^a)_{\a\b}\ve_{ab}\cB^b\cC_i{}^j
~,
\label{Algebra-1.2}
\eea
\bea
{[}\de_a,\de_{\b j}{]}&=&
\Big(
(\g_a)_{\b}{}^{\g}\big(\ri \d_j^k\cS+\cS_{j}{}^{k}\big)
+\ve_{ab}(\g^b)_{\b}{}^{\g}\big(\d_{j}^k\cT+\ri\cT_{j}{}^{k}\big)
+\ri\d_\b^\g\d_j^k\cB_a
+\ri \d_j^k(\g^3)_{\b}{}^{\g}\ve_{ab}\cB^b
\Big)\de_{\g k}
\non\\
&&
+\Big(
C_{jk}\d_{\b}^{\g}A_a
+C_{jk}(\g^3)_{\b}{}^{\g}\ve_{ab}A^b
+C_{jk}\ve_{ab}(\g^b)_{\b}{}^{\g}N
+(\g_a)_{\b}{}^{\g}Y_{jk}
\Big)\deb_\g^k
\non\\
&&
+\Big((\g_a)_{\b}{}^\g\deb_{\g j}N
+{2\over 3}\ve_{ab}(\g^b)_{\b}{}^\g\de_{\g}^k\cS_{jk}
+{2\ri\over 3}(\g_a)_{\b}{}^\g\de_{\g}^{k}\cT_{jk}
-{1\over 3}\ve_{ab}(\g^b)_{\b}{}^\g\deb_{\g}^{k}Y_{jk}
\Big)\cM
\non\\
&&
+\Big(-{1\over 2} \ve_{ab}(\g^b)_{\b}{}^{\g}\d_j^{(k}\deb_\g^{l)}N
+{1\over 6}(\g_a)_{\b}{}^{\g}\d_j^{(k}\de_{\g p}\cS^{l)p}
+{\ri\over 6}\ve_{ab}(\g^b)_{\b}{}^{\g}\d_j^{(k}\de_{\g p}\cT^{l)p}
\non\\
&&~~~
-{1\over 12} (\g_a)_{\b}{}^{\g}\deb_\g^{(k}Y^{lp)}C_{pj}
-{1\over 2} \d_j^{(k}A_a{}_\b^{l)}
\Big)\cV_{kl}
\non\\
&&
+\Big(-{1\over 6}\ve_{ab}(\g^b)_{\b}{}^{\g}\d_j^{(k}\de_{\g p}\cS^{l)p}
-{\ri\over 6}(\g_a)_{\b}{}^{\d}\d_j^{(k}\de_{\d p}\cT^{l)p}
+{1\over 6} \ve_{ab}(\g^b)_{\b}{}^{\d}\d_j^{(k}\deb_{\d p}Y^{l)p}
\non\\
&&~~~
-{1\over 12} \ve_{ab}(\g^b)_{\b}{}^{\d}\deb_\d^{(k}Y^{lp)}C_{pj}
-{1\over 2} \ve_{ab}\d_j^{(k}A^b{}_\b^{l)}
\Big)\cC_{kl}
~,
\label{Algebra-3/2}
\eea
\bea
{[}\de_a,\deb_{\b}^{j}{]}&=&
\Big(
(\g_a)_{\b}{}^{\g}\big(\ri \d^j_k\cS+\cS^{j}{}_{k}\big)
-\ve_{ab}(\g^b)_{\b}{}^{\g}\big(\d^{j}_k\cT+\ri\cT^{j}{}_{k}\big)
-\ri\d_\b^\g\d^j_k\cB_a
-\ri \d^j_k(\g^3)_{\b}{}^{\g}\ve_{ab}\cB^b
\Big)\deb_{\g}^{k}
\non\\
&&
+\Big(
C^{jk}\d_{\b}^{\g}\bar{A}_a
+C^{jk}(\g^3)_{\b}{}^{\g}\ve_{ab}\bar{A}^b
-C^{jk}\ve_{ab}(\g^b)_{\b}{}^{\g}\bar{N}
-(\g_a)_{\b}{}^{\g}\bar{Y}^{jk}
\Big)\de_{\g k}
\non\\
&&
+\Big((\g_a)_{\b}{}^\g\de_{\g}^j\bar{N}
+{2\over 3}\ve_{ab}(\g^b)_{\b}{}^\g\deb_{\g k}\cS^{jk}
+{2\ri\over 3}(\g_a)_{\b}{}^\g\deb_{\g k}\cT^{jk}
+{1\over 3}\ve_{ab}(\g^b)_{\b}{}^\g\de_{\g k}\bar{Y}^{jk}
\Big)\cM
\non\\
&&
+\Big(-{1\over 2} C^{j(k}\ve_{ab}(\g^b)_{\b}{}^{\g}\de_{\g}^{l)}\bar{N}
-{1\over 6}C^{j(k}(\g_a)_{\b}{}^{\g}\deb_{\g p}\cS^{l)p}
+{\ri\over 6}C^{j(k}\ve_{ab}(\g^b)_{\b}{}^{\g}\deb_{\g p}\cT^{l)p}
\non\\
&&~~~
-{1\over 12} (\g_a)_{\b}{}^{\g}\de_{\g}^{(j}\bar{Y}^{kl)}
+{1\over 2} C^{j(k}\bar{A}_a{}_{\b}^{l)}
\Big)\cV_{kl}
\non\\
&&
+\Big({1\over 6}C^{j(k}\ve_{ab}(\g^b)_{\b}{}^{\g}\deb_{\g p}\cS^{l)p}
-{\ri\over 6}C^{j(k}(\g_a)_{\b}{}^{\d}\deb_{\d p}\cT^{l)p}
+{1\over 6} C^{j(k}\ve_{ab}(\g^b)_{\b}{}^{\d}\de_{\d p}\bar{Y}^{l)p}
\non\\
&&~~~
-{1\over 12} \ve_{ab}(\g^b)_{\b}{}^{\d}\de_{\d}^{(j}\bar{Y}^{kl)}
+{1\over 2} C^{j(k}\ve_{ab}\bar{A}^b{}_{\b}^{l)}
\Big)\cC_{kl}
~,
\label{Algebra-3/2c}
\eea
\bea
{[}\de_a,\de_b{]}&=&
-\hf\ve_{ab}\Bigg(
\Big(
\ri\de^{\g k}\bar{N}
+{2\ri\over 3}(\g^3)^{\g\d}\deb_{\d l}\cS^{lk}
+{2\over 3}\deb^{\g}_l\cT^{lk}
+{\ri\over 3}(\g^3)^{\g\d}\de_{\d l}\bar{Y}^{lk}\Big)\de_{\g k}
\non\\
&&
~+\Big(
\ri\deb^\g_{k}N
+{2\ri\over 3}(\g^3)^{\g\d}\de_{\d}^{l}\cS_{lk}
-{2\over 3}\de^{\g l}\cT_{lk}
-{\ri\over 3}(\g^3)^{\g\d}\deb_{\d}^lY_{lk}\Big)\deb_\g^k
\non\\
&&
~+\Big({\ri\over 4}(\g^3)^{\a\b}{[}\deb_{\a k},\deb_{\b}^k{]}N
-{\ri\over 4}(\g^3)^{\a\b}{[}\de_{\a k},\de_{\b}^k{]}\bar{N}
+{\ri\over 12}{[}\deb_{\a k},\deb^\a_{l}{]}Y^{kl}
\non\\
&&
~~~~~~
-{\ri\over 12}{[}\de_{\a k},\de^\a_{l}{]}\bar{Y}^{kl}
-{\ri\over 6}{[}\de_{\a k},\deb^\a_{l}{]}\cS^{kl}
-{1\over 6}(\g^3)^{\a\b}{[}\de_{\a k},\deb_{\b l}{]}\cT^{kl}
\non\\
&&
~~~~~~
+8\cT^2
+4\cT^{kl}\cT_{kl}
+8\cS^2
+4\cS^{kl}\cS_{kl}
+8\bar{N}N
+4\bar{Y}^{kl}Y_{kl}
\Big)\cm
\non\\
&&
~+\Big({\ri\over 16}{[}\de^{\a(k},\de_\a^{l)}{]}\bar{N}
-{\ri\over 16}{[}\deb^{\a(k},\deb_{\a}^{l)}{]}N
-{\ri\over 16}(\g^3)^{\a\b}{[}\deb_{\a p},\deb_{\b}^{p}{]}Y^{kl}
\non\\
&&
~~~~~~
+{\ri\over 16}(\g^3)^{\a\b}{[}\de_{\a p},\de_{\b}^p{]}\bar{Y}^{kl}
+8\cS^{kl}\cT
+8\ri\cS^{(k}{}_p\cT^{l)p}
\Big)\cV_{kl}
\non\\
&&
~+\Big({\ri\over 48}{[}\deb^{\a(k},\deb_{\a p}{]}Y^{l)p}
-{\ri\over 48}{[}\de^{\a(k},\de_{\a p}{]}\bar{Y}^{l)p}
-4\bar{Y}^{p(k}Y_p{}^{l)}\Big)\cC_{kl}
\Bigg)
~.~~~~~~
\label{Algebra-2}
\eea
\end{subequations}
Here the dimension-1 components of the torsion obey the symmetry relations
\bea
Y_{ij}=Y_{ji}~,~~~
\cT_{ij}=\cT_{ji}~,~~~
\cS_{ij}=\cS_{ji}~,
\eea
and the reality conditions
\begin{subequations}
\bea
&(N)^*=\bar{N}~,~~~
(\cT)^*=\cT~,~~~
(\cS)^*=\cS~,~~~
(A_a)^*=\bar{A}_a~,~~~
(\cB_a)^*=\cB_a~,
\\
&(Y^{ij})^*=\bar{Y}_{ij}~,~~~
(\cT^{ij})^*=\cT_{ij}~,~~~
(\cS^{ij})^*=\cS_{ij}~.
\eea
\end{subequations}
All of the previous superfields are Lorentz scalars except the vectors $A_a$ and $\cB_a$.
All of the superfields except $Y_{ij},\,\cS_{ij},\,\cT_{ij}$ are invariant 
under the action of the SU(2)$_\cV$  generator $\cV_{kl}$.  
For $Y_{ij}$ it holds
\bea
\cV_{kl}Y_{ij}=\hf\big(C_{i(k}Y_{l)j}+C_{j(k}Y_{l)i}\big)~,
\eea
with $\cS_{ij},\,\cT_{ij}$ enjoying the same SU(2)$_\cV$ transformation properties of $Y_{ij}$.
The transformation rules of the dimension-1 components of the torsion under the action
of the  $\cC_{kl}$ operator are less trivial. 
It holds\footnote{These relations can be obtained for example by computing 
$[\cC_{kl},\{\de_{\a i},\de_{\b j}\}]$, $[\cC_{kl},\{\de_{\a i},\deb_{\b}^j\}]$ and
$[\cC_{kl},\{\deb_{\a}^{ i},\deb_{\b}^{ j}\}]$ and by using the equations 
(\ref{Algebra-1.1})--(\ref{Algebra-1.2}) together with the commutation relations of the structure 
group operators given in appendix A  eq. (\ref{[VV]})--(\ref{[CC]}).}
\bsubeq
\bea
&\cC_{kl}\cT=-\cS_{kl}~,~~\cC_{kl}\cT_{ij}=-C_{i(k}C_{l)j}\cS~,~~~~
\cC_{kl}\cS=-\cT_{kl}~,~~\cC_{kl}\cS_{ij}=-C_{i(k}C_{l)j}\cT~,~~~~~~
\label{structT1}
\\
&\cC_{kl}N=-Y_{kl}~,~~\cC_{kl}Y_{ij}=-C_{i(k}C_{l)j}N~,~~~~
\cC_{kl}\bar{N}=-\bar{Y}_{kl}~,~~\cC_{kl}\bar{Y}_{ij}=-C_{i(k}C_{l)j}\bar{N}~,~~~~~~
\label{structT2}
\\
&\cC_{kl}A_{a}=\cC_{kl}\bar{A}_{a}=\cC_{kl}\cB_a=0~.
\label{structT3}
\eea
\esubeq

The components of the dimension-1 torsion obey differential constraints imposed by the Bianchi 
identities.
At dimension-3/2 the Bianchi identities give
\bsubeq
\bea
\de_{\a}^iY^{jk}&=&(\g^3)_{\a}{}^{\b}C^{i(j}\de_{\b}^{k)}N~,~~~
\label{3/2-a}
\\
\de_{\a}^iA_b&=&-\ve_{bc}(\g^c)_{\a}{}^{\b}\de_{\b}^iN
~,~~~
\\
\deb_\b^{(j}Y^{kl)}
&=&
-2\de_{\b}^{(j}\cS^{kl)}
=-2\ri(\g^3)_\b{}^{\g}\de_{\g}^{(j}\cT^{kl)}
~,
\\
\de_{\a}^{i}\cS
&=&
{\ri\over 2}(\g^3)_\a{}^{\b}\deb_\b^{i}N
+{1\over 3}(\g^3)_\a{}^{\b}\de_{\b k}\cT^{ki}
-{\ri\over 6}\deb_{\a k}Y^{ki}
~,
\\
\de_{\a}^i\cT
&=&
-\hf\deb_\a^{i}N
+{1\over 3}(\g^3)_\a{}^{\b}\de_{\b k}\cS^{ki}
+{1\over 6}(\g^3)_\a{}^{\b}\deb_{\b k} Y^{ki}
~,
\\
\deb_\b^jA_a&=&A_{a}{}_{\b}^j
-{1\over 3}(\g_a)_\b{}^\g\de_{\g k}\cS^{kj}
+{\ri\over 3}\ve_{ab}(\g^b)_\b{}^\g\de_{\g k}\cT^{kj}
~,~~~
(\g^a)_\a{}^\b A_{a}{}_{\b}^j=0~,~~~
\\
 \de_{\b}^{j}\cB_a
&=&
{\ri\over 6}(\g_a)_{\b}{}^{\g}\de_{\g k}\cS^{jk}
-{1\over 6}\ve_{ad}(\g^d)_{\b}{}^{\g}\de_{\g k}\cT^{jk}
+{\ri\over 6}(\g_a)_{\b}{}^{\g}\deb_{\g k}Y^{jk}
+{\ri\over 2}A_a{}_\b^j
~,~~~~~~~~~
\label{3/2-g}
\eea
\esubeq
where the dimension-3/2 superfield ${A}_a{}_{\b}^{ j}$ 
has been introduced as the gamma-traceless part of $\deb_\b^j{A}_a$
according to (\ref{cA-3/2}).
In the list of dimension-3/2 Bianchi identities we have omitted 
relations that can be easily 
obtained by complex conjugation of (\ref{3/2-a})--(\ref{3/2-g}).

It is worth noting that the supergravity multiplet is completely determined by the previous 
dimension-3/2 differential constraints. The dimension-2 Bianchi identities are solved by making 
use of (\ref{3/2-a})--(\ref{3/2-g}) and differential equations which are consequences of those ones 
(see appendix B.3).

We conclude by noting that if one imposes $Y_{ij}=\cS_{ij}=\cT_{ij}=A_a=\cB_a=0$ the algebra 
and constraints reduce, up to trivial field redefinitions, to the minimal supergravity multiplet of 
Gates {\it et.~al.} \cite{2DN4SG}. 
We will discuss more about the connection between our supergravity formulation and the minimal 
one later in subsection \ref{minimal} and \ref{minimal2}.


\subsection{Super-Weyl transformations}

Here we consider super-Weyl transformations in analogy to the 
analysis of Howe and Tucker \cite{HoweTucker}.
By direct computation, it can be shown that the constraints (\ref{constr-0})--(\ref{constr-1})
are invariant under the finite super-Weyl transformations of the form:
\bsubeq
\bea
 \de'{}_{\a i}&=&\re^{\hf S}\Big(\de_{\a i}
 +(\g^3)_\a{}^\g(\de_{\g i}S)\cM
 -(\de_{\a k}S)\cV_i{}^k
 -(\g^3)_\a{}^\g(\de_{\g k}S)\cC_i{}^k\Big)~,
 \label{SW-1}
\\
\deb'{}_{\a}^i&=&\re^{\hf S}\Big(\deb_{\a}^i
 +(\g^3)_\a{}^\g(\deb_{\g}^iS)\cM
 +(\deb_{\a}^kS)\cV^i{}_k
 +(\g^3)_\a{}^\g(\deb_{\g}^kS)\cC^i{}_k\Big)~,
\label{SW-1c}
\\
\de'{}_a&=&\re^{S}\Big(
\de_a
+{\ri\over 2}(\g_a)^{\g\d}(\de_{\g k}S)\deb_{\d}^k
 -{\ri\over 2}(\g_a)^{\g\d}(\deb_{\g k}S)\de_{\d}^k
+\ve_{ab}(\de^bS)\cM
\non\\
&&~~~
- {\ri\over 8}(\g_a)^{\g\d}({[}\de_{\g}^k,\deb_{\d}^l{]}S)\cV_{kl}
-{\ri\over 8}\ve_{ab}(\g^b)^{\g\d}({[}\de_{\g}^k,\deb_{\d}^l{]}S)\cC_{kl}
\non\\
&&~~~
-{\ri\over 2}\ve_{ab}(\g^b)^{\g\d}(\de_{\g}^kS)(\deb_{\d}^lS)\cC_{kl}
\Big)~.
\label{SW-2}
\eea
\esubeq
Here the parameter $S(z)$ is a real unconstrained superfield $(S)^*=S$
(not to be confused with the torsion component $\cS$).
To ensure the invariance of the algebra under the super-Weyl transformations, the dimension-1 
components of the torsion have to transform as\bsubeq\bea
N'&=&\re^{S}\Big(N+{\ri\over 8} (\g^3)^{\g\d}(\de_{\g k}\de_{\d}^kS)\Big)~,
\label{SW-T}
\\
\cT'&=&\re^S\Big(\cT +{\ri\over 16}(\g^3)^{\g\d}({[}\de_{\g k},\deb_{\d}^k{]}S)\Big)
~,
\\
\cS'&=&\re^S\Big(\cS+{1\over 16}({[}\de_{\g k},\deb^{\g k}{]}S)\Big)
~,
\\
Y'_{ij}&=&\re^{S}\Big(Y_{ij}+{\ri\over 8} (\de_{\g (i}\de_{j)}^\g S)\Big)~,
\label{S_ij-SW}
\\
\cT'_{ij}&=&\re^S\Big(\cT_{ij} +{1\over 16}(\g^3)^{\g\d}({[}\de_{\g (i},\deb_{\d j)}{]}S)\Big)
\label{cT_ij-SW}
~,
\\
\cS'_{ij}&=&\re^S\Big(\cS_{ij}  +{\ri\over 16}({[}\de_{\g (i},\deb^\g_{j)}{]}S)\Big)
\label{cS_ij-SW}
~,
\\
A'_a&=&\re^{S}\Big(A_a
-{\ri\over 8} (\g_a)^{\g\d}(\de_{\g k}\de_{\d}^kS)
-{3\ri\over 8}(\g_a)^{\g\d}(\de_{\g k}S)(\de_{\d}^kS)
\Big)
~,
\label{A_a-SW}
\\
\cB'_a&=&\re^S\Big(\cB_a 
 -{1\over 16}(\g_a)^{\g\d} ({[}\de_{\g k},\deb_{\d}^k{]}S)
 -{3\over 8}(\g_a)^{\g\d}(\de_{\g k}S)(\deb_{\d}^kS)
\Big)
~,~~~
\label{SW-cA}
\eea
\esubeq
together with their complex conjugates.
The proof that the covariant derivatives algebra of subsection \ref{ExtendedSUGRA} 
is invariant under the previous super-Weyl 
transformations is quite long but straightforward and it is left as a useful exercise
to the interested reader.

For later use, we rewrite the super-Weyl transformations of the spinor covariant 
derivatives (\ref{SW-1}) and (\ref{SW-1c}) 
in a form where the left/right Lorentz spinor indices 
are explicit\bsubeq \bea
 \de'{}_{+ i}&=&\re^{\hf S}\Big(\de_{+ i}
 +(\de_{+ i}S)\cM
 +2(\de_{+}^kS)\bmL_{ik}
 \Big)~,
\label{SW+} 
\\
\deb'{}_{+}^i&=&\re^{\hf S}\Big(\deb_{+}^i
 +(\deb_{+}^iS)\cM
 -2(\deb_{+ k}S)\bmL^{ik}
 \Big)~,
 \label{SW+c}
 \\
 \de'{}_{- i}&=&\re^{\hf S}\Big(\de_{- i}
 -(\de_{- i}S)\cM
 +2(\de_{-}^kS)\bmR_{ik}
 \Big)~,
 \label{SW-}
\\
\deb'{}_{-}^i&=&\re^{\hf S}\Big(\deb_{-}^i
 -(\deb_{-}^iS)\cM
 -2(\deb_{- k}S)\bmR^{ik}
 \Big)~.
 \label{SW-c}
\eea
\esubeq
In this form one observes that, under super-Weyl transformations, only the SU(2)$_L$ connections 
of the left covariant spinor derivatives transform  non-homogeneously   and, similarly, 
only the SU(2)$_R$ connections of the right spinor derivatives transform non-homogeneously.

Observing (\ref{SW-T})--(\ref{SW-cA}), it is clear that one can gauge away all 
the theta independent dimension-1 components of the torsion.
In particular using both super-Weyl and the supergravity gauge transformations
one could choose a Wess-Zumino gauge in which the remaining fields are those of the 
Weyl multiplet of conformal supergravity which, in particular,
 does not contain auxiliary fields \cite{2DN4SG}.
One easy way to prove this statement is by noting that under gauge fixing, the extended 
supergravity 
multiplet reduces to the minimal one and then follow the discussion of \cite{2DN4SG}.
More on the fields content and the Wess-Zumino superspace reduction of the 
SO(1,1)$\times$SU(2)$_L\times$SU(2)$_R$ extended geometry is planned to be the subject 
of a separate analysis and is beyond the scope of this paper.

\subsection{On the minimal supergravity multiplet}
\label{minimal}

So far in this section we have introduced a new superspace formulation for
an extended supergravity multiplet having the structure group
SO(1,1)$\times$SU(2)$_L\times$SU(2)$_R$. 
Its super-Weyl transformations, generated by an 
unconstrained real scalar superfield, induce homogeneous transformations on the 
inverse supervielbein in the spinor derivatives (\ref{SW-1}), (\ref{SW-1c}).
We have already mentioned that the extended multiplet
can be gauged fixed to the minimal supergravity multiplet.
For most applications, the minimal formulation is more convenient to work with 
even if, as explicitly described in the following, the super-Weyl 
transformations are more tricky.
Let us consider here in greater detail the implications of the minimal gauge fixing.

First, we impose the following gauge condition in the supergravity multiplet
\bea
\cS_{ij}=\cT_{ij}=Y_{ij}=A_a=\cB_a=0~.
\label{minimalGAUGE}
\eea
It can  proved that the superfields 
$\cS_{ij},\,\cT_{ij},\,Y_{ij},\,A_a$ and $\cB_a$
are pure gauge degrees of freedom under super-Weyl 
transformations; we will come back to
this important point in subsection \ref{minimal2}.

One readily observes that under (\ref{minimalGAUGE}) 
all the $R_\cC$ curvatures are identically zero 
and we can choose 
\bea
(\F_\cC)_A{}^{kl}=0~,
\label{minimalGAUGE222}
\eea
 in the covariant derivatives (\ref{CovDev-4}).
The resulting constraints on the surviving superfields $N,\,\cS,\cT$ are 
\bea
\de_{\b}^{j}N=0~,~~~
\de_{\a}^{i}\cS
=
{\ri\over 2}(\g^3)_\a{}^{\b}\deb_\b^{i}N
~,~~~
\de_{\a}^i\cT
=
-\hf\deb_\a^{i}N
~.
\eea
These, up to field redefinitions, are the constraints that characterize the dimension-1
torsion components of the minimal supergravity of \cite{2DN4SG}.
In particular they describe a covariant extension of the dimension-1/2 differential constraints of 
the twisted-I multiplet \cite{Gnew2D,GHR,GatesKetov}.

The structure group of the resulting minimal multiplet 
now has a remaining local SO(1,1)$\times$SU(2)$_\cV$ symmetry. Moreover,
the gauge choice (\ref{minimalGAUGE}) still has residual super-Weyl transformations
(\ref{SW-T})--(\ref{SW-cA}).
For simplicity, we restrict ourselves to infinitesimal transformations; the finite transformations 
can be easily derived along the same lines.
To distinguish between the super-Weyl transformations of the extended and minimal
geometry, we redefine in the minimal case the real superfield $S$ with $\bS$.
Let us look again at the transformation (\ref{SW-1}), which in the infinitesimal limit is
\bea
\d \de_{\a i}&=&
\hf\bS\de_{\a i}
 +(\g^3)_\a{}^\g(\de_{\g i}\bS)\cM
 -(\de_{\a k}\bS)\cV_i{}^k
 -(\g^3)_\a{}^\g(\de_{\g k}\bS)\cC_i{}^k
~.
\label{infSW}
\eea
The last term in (\ref{infSW}) tells us that the super-Weyl transformations alone break
the gauge $(\F_\cC)_A{}^{kl}=0$. This can be fixed by adding a compensating SU(2)$_\cC$ 
transformation to cancel the induced $(\F_\cC)_{\a i}{}^{kl}$ spinor connection in (\ref{infSW}).

An infinitesimal SU(2)$_\cC$ transformation of the spinor covariant derivatives, with real 
parameter $\bS_{ij}=(\bS^{ij})^*$, is 
\bea
\d_\cC\de_{\a i}=[\bS^{kl}\cC_{kl},\de_{\a i}]
=
-(\g^3)_\a{}^\b\bS_i{}^{j}\de_{\b j}
-(\de_{\a i}\bS_{kl})\cC^{kl}
~.
\eea
Imposing the following differential constraint between $\bS$ and $\bS_{ij}$
\bea
(\de_{\a i}\bS_{kl})=-\hf(\g^3)_\a{}^\b C_{i(k}(\de_{\b k)}\bS)~,
\label{dbSbS}
\eea
one obtains the modified super-Weyl transformation that preserves the gauge 
$(\F_\cC)_A{}^{kl}=0$.
This is given by $\tilde{\d}=(\d-\d_\cC)$ 
\bea
\tilde{\d} \de_{\a i}&=&
\hf\bS\de_{\a i}
+(\g^3)_\a{}^\b\bS_i{}^{j}\de_{\b j}
 +(\g^3)_\a{}^\g(\de_{\g i}\bS)\cM
 -(\de_{\a k}\bS)\cV_i{}^k
~.
\label{infSW-2}
\eea
Note that due to the compensating SU(2)$_\cC$ transformation, the supervielbein 
in (\ref{infSW-2}) does not 
transform homogeneously anymore.
Equation (\ref{infSW-2}) was first derived in \cite{2DN4SG}.

Note that eq. (\ref{dbSbS}) is the dimension-1/2 differential constraint of a twisted-II multiplet 
\cite{TM-II-IK,GatesKetov}.
It implies the following dimension-1 differential constraints on $\bS$ and $\bS_{ij}$\bsubeq\bea
\de_{\a i}\de_{\b j}\bS&=&
-4\ri C_{\a\b}N\bS_{ij}
+{1\over 6}C_{ij}(\g^3)_{\a\b} (\de_{\g}^k\de^{\g l}\bS_{kl})
~,
\\
{[}\de_\a^i,\deb_{\b}^{j}{]}\bS&=&
-{1\over 6}C^{ij}C_{\a\b}(\g^3)^{\g\d}[\de_{\g k},\deb_{\d l}]\bS^{kl}
+{1\over 6}C^{ij}(\g^3)_{\a\b}[\de_{\d k},\deb_{l}^\d]\bS^{kl}
\non\\
&&
+4\ri\ve^{ab}(\g_a)_{\a\b}\de_b\bS^{ij}
-8\big((\g^3)_{\a\b}\cS+\ri C_{\a\b}\cT\big)\bS^{ij}
~.
\label{ddbS}
\eea
\esubeq
By using the previous two results and (\ref{structT1})--(\ref{structT3}), it can be explicitly observed
that (\ref{minimalGAUGE}) are preserved by the $\tilde{\d}$ transformation
and that the dimension-1 torsion components of the minimal multiplet transform according to
 the following rules
\bsubeq
\bea
\tilde{\d} N&=&\bS N+{\ri\over 8} (\g^3)^{\g\d}(\de_{\g k}\de_{\d}^k\bS)
~,
\label{minimalSW-1}
\\
\tilde{\d}\cT&=&\bS\cT +{\ri\over 16}(\g^3)^{\g\d}({[}\de_{\g k},\deb_{\d}^k{]}\bS)
~,
\label{minimalSW-2}
\\
\tilde{\d}\cS&=&\bS\cS+{1\over 16}({[}\de_{\g k},\deb^{\g k}{]}\bS)
~.
\label{minimalSW-3}
\eea
\esubeq
The transformations of the $\deb_\a^i$ covariant derivative can be trivially obtained by 
complex conjugation of (\ref{infSW-2}). 
We conclude by observing that for the vector covariant derivative it holds that
\bea
\tilde{\d}\de_a&=&
\bS\de_a
+{\ri\over 2}(\g_a)^{\g\d}(\de_{\g k}\bS)\deb_{\d}^k
 +{\ri\over 2}(\g_a)^{\g\d}(\deb_{\g}^{k}\bS)\de_{\d k}
 \non\\
 &&
+\ve_{ab}(\de^b\bS)\cM
- \ve_{ab}(\de^b\bS^{kl})\cV_{kl}
~,~~~~~~~~~
\eea
where (\ref{ddbS}) has been used.


\section{Coupling to an Abelian vector multiplet}
\setcounter{equation}{0}
\label{VectorMultiplet}

Let us couple the extended conformal supergravity multiplet to an off-shell vector multiplet.
We describe here in detail the case of a single Abelian vector multiplet, which will be
interpreted as a real central charge. 
The resulting multiplet is of particular importance since it plays the role of a conformal 
compensator for supergravity.
The covariant vector multiplet has field strength described by a scalar twisted-II multiplet.
The covariant coupling with the algebra is useful because the structure group  and super-Weyl 
transformations will be easily indicated by consistency of the geometry.

\subsection{Twisted-II vector multiplet}

The coupling of the supergravity geometry to an Abelian vector multiplet
is achieved by modifying the covariant derivatives as follows
\bea
{\bm \de}_{A}&=&
\de_A
+V_A{\bm Z}
~,\label{CovDev-2-2}
\eea
with $V_A(z)$ the U(1)$_Z$ gauge connection.
The gauge transformations of the covariant derivatives are
\bea
\d_{Z}\de_A=[\tau{\bm Z},\de_A]~,
\eea
with $\tau(z)$ the parameter of the U(1)$_Z$ transformations.
The operator ${\bm Z}$ is conveniently interpreted as a real central charge 
$({\bm Z})^*={\bm Z}$.
The multiplet introduced in this way is reducible. One can then impose
appropriate covariant constraints on some components of the gauge-invariant field strength
$F_{AB}$  which appears in the algebra of gauge-covariant derivatives
\bea
{[}{\bm \de}_{{A}},{\bm \de}_{{B}}\}&=&T_{{A}{B}}{}^{{C}}{\bm \de}_{{C}}
+R_{{A}{B}}\cM
+(R_\cV)_{{A}{B}}{}^{kl}\cV_{kl}
+(R_\cC)_{{A}{B}}{}^{kl}\cC_{kl}
+F_{AB}{\bm Z}
~.
\label{algebra-2-2}
\eea
For consistency the field strength $F_{AB}$ has to satisfy the Bianchi identities
\bea
\sum_{[ABC)}\Big(\de_{{A}}F_{{B}{C}}-T_{{A}{B}}{}^{{D}}F_{{D}{C}}\Big)=0~.
\label{Cz-Bianchi04}
\eea
Here a graded cyclic sum was assumed.
The torsion $T_{AB}{}^C$ and curvatures 
$R_{{A}{B}},\,(R_\cV)_{{A}{B}}{}^{kl}$ and $(R_\cC)_{{A}{B}}{}^{kl}$ are the ones appearing
in (\ref{Algebra-1.1})--(\ref{Algebra-2}).
Note that in (\ref{Cz-Bianchi04}) we used the $\de_A$ derivatives instead 
of ${\bm \de}_A$ since the field strength is neutral with respect to the central charge ${\bm Z}$.
Since the torsion and curvatures are also neutral, we will always use $\de_A$ in the Bianchi 
identities.

In the limit of flat superspace one can easily find two distinct irreducible representations for the 
vector multiplet field strength \cite{Siegel2DN4}.\footnote{In 2D $\cN=(2,2)$, dual formulations of 
minimal vector multiplets are also known, {\it{e.~g.}}  \cite{GatesMerrell,LRRvUZ}.}
The first is described by the constraints\begin{subequations}\bea
&&F_{\a i}{}_{\b j}=-2C_{\a\b}C_{ij}\bar{W}~,~~
F_{\a}^i{}_{\b}^j=-2C_{\a\b}C^{ij}W~,~~
F_{\a i}{}_\b^j=2\ri\d_i^j\big(C_{\a\b}P+\ri(\g^3)_{\a\b}Q\big)~,
~~~~~~
\label{C-TMI-1}
\\
&&
F_a{}_{\b j}=-{\ri\over 2}(\g_a)_{\b}{}^\g\DB_{\g j}\bar{W}
~,~~~
F_a{}_\b^j={\ri\over 2}(\g_a)_{\b}{}^\g D_{\g}^jW~,
\label{C-TMI-2}
\\
&&
F_{ab}=-{1\over 16}\ve_{ab}\Big((\g^3)^{\g\d}D_{\g k}D_{\d}^kW
+(\g^3)^{\g\d}\DB_{\g k}\DB_{\d}^k\bar{W}\Big)~,
\label{C-TMI-3}
\eea
\end{subequations}
where $D_A$ are the flat superspace covariant derivatives.
The complex superfield $W$ ($\bar{W}=(W)^*$) and the real superfields $P,\,Q$
($(P)^*=P,\,(Q)^*=Q$) satisfy the differential constraints of a twisted-I  multiplet (TM-I)
\cite{Gnew2D,GHR,GatesKetov} 
\bea
D_{\a i}\bar{W}=0~,~~\DB_\a^iW=0~,~~~
D_{\a i}P=-{\ri\over 2}\DB_{\a i}\bar{W}~,~~~
D_{\a i}Q=\hf(\g^3)_\a{}^\b\DB_{\b i}\bar{W}~.
\eea
Note that the previous vector multiplet can be easily obtained by dimensionally reducing from 4D 
to 2D the well known 4D, $\cN=2$ vector multiplet constraints \cite{GSW}.

A second irreducible set of constraints for the vector multiplet field strength can be proven 
to be\begin{subequations}\bea
&&F_{\a i}{}_{\b j}=\Big((\g^3)_{\a\b}W_{ij}+\hf C_{\a\b}C_{ij}F\Big)~,~
F_{\a}^i{}_{\b}^j=\Big((\g^3)_{\a\b}W^{ij}+\hf C_{\a\b}C^{ij}F\Big)~,~
\label{C0-TMII-1}
\\
&&
F_{\a i}{}_\b^j=0~,~~~~~~
F_a{}_{\b j}={\ri\over 4}(\g_a)_{\b}{}^\g\DB_{\g j}F
~,~~~
F_a{}_\b^j=-{\ri\over 4}(\g_a)_{\b}{}^\g D_{\g}^jF~,
\label{C0-TMII-2}
\\
&&
F_{ab}=-{1\over 48}\ve_{ab}\Big(D_{\g}^{k}D^{\g l}W_{kl}
+\DB_{\g}^k\DB^{\g l}W_{kl}\Big)~,
\label{C0-TMII-3}
\eea
\end{subequations}
provided that the real superfields $W_{ij},\,F$ satisfy the constraints
\bea
D_{\a i} W_{jk}+{1\over 2}C_{i(j}(\g^3)_\a{}^\b D_{\b k)}F=0~,~~~~~~
(W_{ij})^*={W}^{ij}~,~~(F)^*=F~.
\eea
Then one sees that $W_{ij}$ and $F$ describe a twisted-II multiplet
\cite{TM-II-IK,GatesKetov}.

It is interesting to note that the previous flat vector multiplet constraints can not be both
consistently lifted to a coupling with the supergravity of subsection \ref{ExtendedSUGRA}.
The point is that once the vector multiplet is coupled to supergravity by using
eq. (\ref{CovDev-2-2}) and (\ref{algebra-2-2}), 
the structure group and super-Weyl transformation properties 
of the vector multiplet field strength $F_{AB}$ are fixed by the geometry.
In particular, by considering the commutator $[\cC_{kl},\{{\bm \de}_{\a i},{\bm \de}_{\b j}\}]$
and eq. (\ref{algebra-2-2}) together with the constraints (\ref{C-TMI-1}) one observes that 
the TM-I type of constraints on the field strength is inconsistent with the $\cC_{kl}$ transformations.
Therefore, the constraints (\ref{C-TMI-1}) could not be extended 
to our supergravity case without fixing the 
SU(2)$_\cC$ group.\footnote{In the case of minimal supergravity \cite{2DN4SG} one can 
prove that the constraints (\ref{C-TMI-1}) can be consistently coupled to the algebra. 
However, such coupling results to be inconsistent with the super-Weyl transformations of the 
minimal multiplet \cite{GTMunp-1}.}

On the other hand, by using the same arguments, it follows that the constraints (\ref{C0-TMII-1}) 
are consistent with the $\cC_{kl}$ transformations  provided that 
the real superfields $W_{ij},\,F$ satisfy in the curved case 
\bea
\cC_{kl}F=2W_{kl}~,~~~
\cC_{kl}W_{ij}=\hf C_{i(k}C_{l)j}F
~.
\label{C_kl-W_ij-F}
\eea
Then, one can check that the constraints (\ref{C0-TMII-1})--(\ref{C0-TMII-3}),
 in the curved geometry of subsection \ref{ExtendedSUGRA}, become
\begin{subequations}
\bea
&&F_{\a i}{}_{\b j}=\Big((\g^3)_{\a\b}W_{ij}+\hf C_{\a\b}C_{ij}F\Big)~,~
F_{\a}^i{}_{\b}^j=\Big((\g^3)_{\a\b}W^{ij}+\hf C_{\a\b}C^{ij}F\Big)~,~
\\
&&
F_{\a i}{}_\b^j=0~,~~~~~~
F_a{}_{\b j}={\ri\over 4}(\g_a)_{\b}{}^\g\deb_{\g j}F
~,~~~
F_a{}_\b^j=-{\ri\over 4}(\g_a)_{\b}{}^\g \de_{\g}^jF~,
\\
&&
F_{ab}=-{1\over 48}\ve_{ab}\Big(\de_{\g}^{k}\de^{\g l}W_{kl}
+\deb_{\g}^k\deb^{\g l}W_{kl}
+24\ri\big(\bar{N}-N\big)F
+24\ri\big(Y^{kl}-\bar{Y}^{kl}\big)W_{kl}
\Big)~.~~~~~~~~
\eea
\end{subequations}
Here the superfields $W_{ij},\,F$ enjoy the covariant extension of the TM-II differential 
constraints
\bea
\de_{\a i} W_{jk}=-{1\over 2}C_{i(j}(\g^3)_\a{}^\b\de_{\b k)}F~,~~~~~~
(W_{ij})^*={W}^{ij}~,~~~(F)^*=F~,
\label{constr-TMII}
\eea
along with a complex conjugate constraint.
The $W_{ij},\,F$ superfields are Lorentz scalars.
Under SU(2)$_\cV$ transformations it holds
$\cV_{kl} F=0$ and $\cV_{kl}W_{ij}=\hf (C_{i(k}W_{l)j}+C_{j(k}W_{l)i})$.
The SU(2)$_\cC$ transformations are given in (\ref{C_kl-W_ij-F}).
By direct, but not short,
 computations one can prove that the Bianchi identities (\ref{Cz-Bianchi04}) 
are then identically satisfied.

As a final remark we observe that 
the consistency of (\ref{algebra-2-2}) requires the superfields $W_{ij},F$
to transform homogeneously under the super-Weyl transformations 
(\ref{SW-1})--(\ref{SW-2}), i. e.
\bea
W'{}_{ij}=\re^{S}W_{ij}~,~~~F'=\re^{S}F~.
\label{SW-TMII}
\eea
Note that the TM-II differential constraint in eq. (\ref{constr-TMII}), 
is then invariant under super-Weyl transformations (\ref{SW-TMII}).

If one reduces the curved geometry to the one of the
 minimal supergravity multiplet, according to the 
discussion in subsection \ref{minimal},
the consistent infinitesimal super-Weyl transformations are 
\bea
\tilde{\d}W_{ij}=\bS W_{ij}+\bS_{ij}F~,~~~
\tilde{\d}F=\bS F-2\bS^{kl}W_{kl}~.
\eea

\subsection{Chiral prepotential of TM-II and covariant matter TM-I}
\label{subCTMII}

Here we want to prove the following statement:
given a chiral superfield $W$ invariant under structure group and super-Weyl transformations 
\bea
\cm W=\cV_{kl}W=\cC_{kl}W=0~,~~~~~~
W'=W~,
\label{TMII-prep-1}
\eea
and subject to the conditions
\begin{subequations}
\bea
&\deb_\a^iW=0~,~~\de_{\a i}\bar{W}=0~,~~~(W)^*=\bar{W}~,
\label{TMII-prep-2}
\\
&\de_{\a (i}\de^\a_{j)}W
=\deb_{\a (i}\deb^\a_{j)} \bar{W}
~,~~~
(\g^3)^{\a\b}\de_{\a i}\de_{\b}^iW
=(\g^3)^{\a\b}\deb_{\a i}\deb_{\b}^i\bar{W}
~,
\label{TMII-prep-3}
\eea
\end{subequations}
then the real descendant operators defined by
\begin{subequations}
\bea
&&\S_{ij}={1\over 4} \de_{\a i}\de^\a_{j}W
={1\over 4}\deb_{\a i}\deb^\a_{j}\bar{W}=(\S^{ij})^*~,
\label{descend-1}\\
&&\S=-{1\over 4}(\g^3)^{\a\b}\de_{\a i}\de_{\b}^iW
=-{1\over 4}(\g^3)^{\a\b}\deb_{\a i}\deb_{\b}^i\bar{W}=(\S)^*
~,
\label{descend-2}
\eea
\end{subequations}
define a covariant TM-II satisfying all the conditions (\ref{C_kl-W_ij-F}), (\ref{constr-TMII}) and 
(\ref{SW-TMII}) with the identifications $W_{ij}=\S_{ij}$ and $F=\S$.
Alternatively, this states that given a covariant TM-II, 
a constrained prepotential\footnote{It is worth noting that in the flat case a more complete analysis
of TM-II constraints in terms of prepotentials has been described in \cite{Siegel2DN4}.
This partly involved the use of a form of bi-projective superspace.
Within the scope of the present paper the constrained 
prepotentials given in this subsection are enough.}
 is given by a  superfield $W$ satisfying (\ref{TMII-prep-1})--(\ref{TMII-prep-3}).

The proof of the previous statement involves some easy but instructive computations.
Using (\ref{descend-1})--(\ref{descend-2}), (\ref{TMII-prep-1}),
(\ref{Lorentz1})--(\ref{C_kl}) one obtains $\cm\S=\cm\S_{ij}=\cV_{kl}\S=0$ and
\bea
\cV_{kl}\S_{ij}=\hf\big(C_{i(k}\S_{l)j}+C_{j(k}\S_{l)i}\big)~,~~~
\cC_{kl}\S=2\S_{kl}~,~~~
\cC_{kl}\S_{ij}=\hf C_{i(k}C_{l)j}\S
~.
\eea
Some $\de$-algebra gives
\bsubeq
\bea
&&~~~~~~~~~~~~~~~~~~~~~
\de_{\a(i}\S_{jk)}=\deb_{\a(i}\S_{jk)}=0~,
\label{S111}
\\
&&
\de_\a^j\S_{ij}
=
{3\ri\over 2} (\g^a)_{\a}{}^{\b}\de_a\deb_{\b i}\bar{W}
~,
~~~~~~
\de_{\a i}\S
=
\ri\ve^{ab}(\g_a)_\a{}^\b\de_b\deb_{\b i}\bar{W}
~.
\label{S113}
\eea
\esubeq
The equations (\ref{S111})--(\ref{S113}) then imply the TM-II differential constraint 
(\ref{constr-TMII})
\bea
\de_{\a i} \S_{jk}&=&-{1\over 2}C_{i(j}(\g^3)_\a{}^\b\de_{\b k)}\S~.
\eea
To conclude the proof that $\S_{ij},\,\S$ describe a TM-II according to (\ref{C_kl-W_ij-F}), 
(\ref{constr-TMII}) and (\ref{SW-TMII}), one has to prove that under super-Weyl transformations
it holds $(\S_{ij})'=\re^{S}\S_{ij}$ and $\S'=\re^{S}\S$. This can be easily seen by using
the equations (\ref{TMII-prep-1})--(\ref{descend-2}) and the super-Weyl transformations of the 
covariant derivatives (\ref{SW-1}) and (\ref{SW-1c}).

An irreducible realization for the superfield $W$ is given by the chiral component of a 
covariant twisted-I multiplet.
This is described by the superfields $W,\,P$ and $Q$.
They are consistently chosen to be invariant under all the
SO(1,1)$\times$SU(2)$_L\times$SU(2)$_R$ and super-Weyl transformations and enjoy 
the following constraints\footnote{The invariance of $W,\,P$ and $Q$ under the structure
group and super-Weyl transformations clearly tells us that this version of the covariant TM-I
can not be embedded in the field strengths of a vector multiplet differently to the covariant 
TM-II considered in this subsection.}
\bsubeq
\bea
&&\deb_\ad^iW=0~,~~~
\de_{\g k}Q=\hf(\g^3)_\g{}^\d\deb_{\d k}\bar{W}~,~~~
\de_{\a i}P=-{\ri\over 2}\deb_{\a i}\bar{W}~,
\label{cov-TM-I-1}
\\
&&~~~~~~~~~~~~~~~~~~
(W)^*=\bar{W}~,~~~(P)^*=P~,~~~(Q)^*=Q~.
\eea
\esubeq
In (\ref{cov-TM-I-1}) we have omitted some constraints that can be obtained by complex 
conjugation.
Using (\ref{cov-TM-I-1}), it is easy to prove the relation
\bea
\de_{\a i}\de_{\b j}W
&=&\deb_{\a i}\deb_{\b j}\bar{W}-4C_{ij}(\g^a)_{\a\b}\de_aP~,
\eea
which implies (\ref{TMII-prep-2}).
It is worth to mention that in \cite{GTMreview2DN4}, where the interested reader is referred,
 we present the solution of the covariant TM-I constraints 
in the language of the bi-projective superspace of section \ref{projSuper}.

We conclude this subsection by remarking that there is a crucial difference
between the TM-I prepotential introduced here
and the supergravity multiplet in the minimal gauge 
(\ref{minimalGAUGE})--(\ref{minimalGAUGE222}) 
described by the torsion components $N,\,\cS$ and $\cT$.
The superfields $(W,\,P,\,Q)$ are invariant under super-Weyl transformations while 
$(N,\,\cS,\,\cT)$
are not and transform inhomogeneously according to (\ref{minimalSW-1})--(\ref{minimalSW-3}).
This difference emphasizes that, even if both the sets of superfields consistently 
satisfy the covariant extensions of the dimension-1/2 TM-I differential constraints,
$(W,\,P,\,Q)$ are matter superfields while $(N,\,\cS,\,\cT)$ are supregravity torsion components.

\subsection{On the minimal supergravity multiplet: II}
\label{minimal2}

In subsection \ref{minimal} we have described the relation between the extended 
SU(2)$_L\times$SU(2)$_R$ supergravity formulation and the minimal SU(2)$_\cV$ multiplet
of \cite{2DN4SG}. Here, by making use of the covariant TM-II multiplet,
 we follow an analogue of the Howe's procedure for 4D $\cN=2$ \cite{Howe}
to introduce the minimal multiplet. 
The analysis goes along the same lines of the 4D $\cN=2$ case described in
\cite{KLRT-M_4D-1,KLRT-M_4D-2}. The 
SU(2)$_L\times$SU(2)$_R$ supergravity multiplet plays the role of the U(2)-Howe
formulation of the 
4D $\cN=2$ Weyl multiplet \cite{Howe}, 
while the SU(2)-Grimm formulation \cite{Grimm} 
is the analogue of the  2D $\cN=(4,4)$ SU(2)$_\cV$ minimal supergravity of \cite{2DN4SG}.

Suppose to have coupled the 2D $\cN=(4,4)$ extended 
supergravity geometry of subsection \ref{ExtendedSUGRA} to a TM-II Abelian vector multiplet
such that at each point of the superspace: (i) $F\ne0$ and, (ii) $W_{ij}=0$.
The second condition can  always  be achieved by the aid of a local
SU(2)$_L\times$SU(2)$_R$ transformation.\footnote{By using
 (\ref{C_kl-W_ij-F}) one observes that the SU(2)$_\cC$
transformation with gauge paramenter $(K_{\cC})_{kl}=(1/F)W_{kl}$ cancels $W_{ij}$ at the 
linearized level; it is not difficult to compute the finite analogue of this result.}
Note that the previous condition is left invariant by SU(2)$_\cV$ transformations but breaks
SU(2)$_\cC$.
Under a super-Weyl transformation (\ref{SW-TMII}) with parameter $S=-\log{F}$
we can then impose the gauge
\bea
F=1~,~~~W_{ij}=0~,
\label{MINIMAL-g}
\eea
which completely fixes the super-Weyl and local SU(2)$_{\cC}$ transformations.

The previous gauge implies various conditions.
First, the covariant constraint (\ref{constr-TMII})
$\de_{\a i} W_{jk}=-{1\over 2}C_{i(j}(\g^3)_\a{}^\b\de_{\b k)}F$, 
in the limit (\ref{MINIMAL-g}), is
\bea
-(\F_\cC)_{\a i}{}_{jk}=0~,
\eea
and therefore the spinor SU(2)$_\cC$ connections are zero
\bea
(\F_\cC)_{\a i}{}^{kl}=(\F_\cC)_{\a}^{i}{}^{kl}=0~.
\eea
Note that due to the previous equations, it follows that in the gauge (\ref{MINIMAL-g})
it also holds the covariantly constant conditions
 $\de_{\a i}F=\deb_\a^iF=\de_{\a i}W_{jk}=\deb_\a^iW_{jk}=0$.

The constraint  (\ref{constr-TMII}) implies in general  the following equations
\bsubeq
\bea
{[}\de_{\a i},\deb_{\b j}{]}F&=&
-{1\over 6}C_{ij}C_{\a\b}(\g^3)^{\g\d} {[}\de_{\g}^k,\deb_{\d}^l{]}W_{kl}
+{1\over 6}C_{ij}(\g^3)_{\a\b} {[}\de_\d^k,\deb^{\d l}{]}W_{kl}
\non\\
&&
+4\ri \ve^{ab}(\g_a)_{\a\b}\de_bW_{ij}
-8\ri C_{\a\b}\cT W_{ij}
-8(\g^3)_{\a\b}\cS W_{ij}
\non\\
&&
-4(\g^3)_{\a\b}\cT_{ij}F
-4\ri C_{\a\b}\cS_{ij}F
+4C_{ij}(\g^a)_{\a\b}\cB_aF
~,
\label{2-111}
\\
\de_{\a i}\de_{\b}^kW_{jk}&=&
-{1\over 4}C_{ij}C_{\a\b}\de_{\g}^k\de^{\g l}W_{kl}
+6\ri(\g^3)_{\a\b}NW_{ij}
-6\ri(\g^a)_{\a\b}A_aW_{ij}
\non\\
&&
+3\ri (\g^3)_{\a\b}Y_{ij}F
-3\ri C_{\a\b}Y_{(i}{}^{p}W_{j)p}
-3\ri C_{ij}(\g_a)_{\a\b}\ve^{ab}A_bF
~.
\label{2-112}
\eea
\esubeq
Equations (\ref{2-111}) and (\ref{2-112}) in the gauge (\ref{MINIMAL-g}) reduce to
\bsubeq
\bea
0&=&
-4\ri \ve^{ab}(\g_a)_{\a\b}(\F_\cC)_b{}_{ij}
-4(\g^3)_{\a\b}\cT_{ij}
-4\ri C_{\a\b}\cS_{ij}
+4C_{ij}(\g^a)_{\a\b}\cB_a
~,
\\
0&=&
3\ri (\g^3)_{\a\b}Y_{ij}
-3\ri C_{ij}(\g_a)_{\a\b}\ve^{ab}A_b
~,
\eea
\esubeq
that imply 
\bsubeq
\bea
&(\F_\cC)_a{}^{kl}=0~,
\\
&\cS_{ij}=\cT_{ij}=Y_{ij}=A_a=\cB_a=0~.
\label{2-2-2}
\eea
\esubeq
It is clear that the gauge (\ref{MINIMAL-g}) reduces the extended supergravity of 
section \ref{ExtendedSUGRA} to the minimal 
multiplet \cite{2DN4SG} of subsection \ref{minimal} coupled to a real constant central charge.

Now, recall the super-Weyl transformations of
 $\cS_{ij},\,\cT_{ij},\,Y_{ij},\,A_a$ and $\cB_a$ eqs. (\ref{S_ij-SW})--(\ref{SW-cA}).
The fact that there exists a gauge in which $\cS'_{ij}=\cT'_{ij}=Y'_{ij}=A'_a=\cB'_a=0$
is equivalent to setting the left hand side of eqs. (\ref{S_ij-SW})--(\ref{SW-cA}) to zero.
This implies that one can solve the
differential constraints of the $\cS_{ij},\,\cT_{ij},\,Y_{ij},\,A_a$ and $\cB_a$
superfields in terms of some real scalar superfields  through the right hand side of eqs. 
(\ref{S_ij-SW})--(\ref{SW-cA}).
Then, we can reinterpret the derivation of eq. (\ref{2-2-2})
in the gauge (\ref{MINIMAL-g}), 
 as a proof that $\cS_{ij},\,\cT_{ij},\,Y_{ij},\,A_a$ and $\cB_a$ are pure gauge degrees of freedom
 where the vector multiplet plays the role of a useful technical tool.
Therefore, the $\cS_{ij},\,\cT_{ij},\,Y_{ij},\,A_a$ and $\cB_a$ superfields, 
in the general case of subsection \ref{ExtendedSUGRA},
 can be gauged away by a super-Weyl transformation.
The previous analysis justify the gauge condition (\ref{minimalGAUGE}) and the results of 
subsection \ref{minimal}.

Note that the previous discussion is similar to the proof we gave in 
\cite{KLRT-M_4D-2} that for the Howe's formulation of 4D $\cN=2$ supergravity
the $G_a{}^{jk}$ superfield is a pure gauge degree of freedom.


\section{2D $\cN=(4,4)$ curved bi-projective superspace}
\setcounter{equation}{0}
\label{projSuper}

In five \cite{KT-M_5D,KT-M_5Dconf} and four \cite{KLRT-M_4D-1,KLRT-M_4D-2} dimensions,
matter couplings in supergravity has been described in terms of covariant projective
supermultiplets. In this section, we introduce the concept of covariant bi-projective
supermultiplets for 2D $\cN=(4,4)$ conformal supergravity, and then
we present a locally supersymmetric and super-Weyl invariant action.
The covariant bi-projective multiplets are a curved extensions of the multiplets introduced in 
the case of 2D $\cN=(4,4)$ flat superspace \cite{BusLinRoc,GHR,RSS,LR-biProj}.
First, let us consider again the TM-II.

Before turning to the details let us make a note for the reader about our notations in this section.
In the sections \ref{SUGRA} and \ref{VectorMultiplet}, 
we have always made use of SU(2)-indices denoted by lower-case letters like $i,\,j$.
In this section we often make use of  
lower-case and capital SU(2)-indices, like $i$ and $I$, to distinguish between indices transforming
respectively only under the SU(2)$_L$ and SU(2)$_R$ group.
For example, according to such distinction, in this section 
we denote the left covariant derivatives as $(\de_{+ i},\,\deb_+^i)$
and the right covariant derivatives as $(\de_{- I},\,\deb_-^I)$.
In using, as in sections \ref{SUGRA} and \ref{VectorMultiplet}, 
the SU(2)$_\cV\times$SU(2)$_\cC$ parametrization of SU(2)$_L\times$SU(2)$_R$,
this index difference is not natural but it turns out to be useful in working with the light-cone 
coordinates.

\subsection{Rewriting the twisted-II multiplet}
\label{TMII-2}

Here we want to give an equivalent description of the twisted-II multiplet in terms of a single
superfield $T_{iI}$ satisfying a set of analyticity-like differential constraints.
This description results to be a covariant extension of the TM-II as introduced for the first time in
the flat superspace case in \cite{TM-II-IK}.

First let us rewrite the TM-II differential constraints (\ref{constr-TMII})  as
\bea
\de_{+ i} W_{jk}=-{1\over 2}C_{i(j}\de_{+ k)}F~,~~~
\de_{- i} W_{jk}={1\over 2}C_{i(j}\de_{- k)}F~,
\label{+-WF}
\eea
where we have explicitly distinguished the left and right Lorentz spinor indices.
We then define the real superfield $T_{iI}$ in terms of $W_{ij}$ and $F$ as
\bea
T_{iI}:=W_{iI}+\hf C_{iI}F~,~~~~~~(T_{iI})^*=T^{iI}~.
\eea
With the previous definition the TM-II differential constraints (\ref{+-WF}) are equivalent to the 
analyticity like constraints
\bea
\de_{+ (i}T_{j)I}=\deb_{+(i}T_{j)I}=0~,~~~
\de_{-(I}T_{|i|J)}=\deb_{-(I}T_{|i|J)}=0
~.
\label{TMII-analyticity}
\eea
The  Lorentz scalar superfield $T_{iI}$ has transformations under the SU(2) groups
defined by the one of $W_{ij},\, F$.
One finds
\bea
\bmL_{kl}T_{iI}=\hf C_{i(k}T_{l)I}~,~~~~~~
\bmR_{KL}T_{iI}=\hf C_{I(K}T_{|i|L)}~.
\label{LTRT}
\eea
Then, it is clear that the index $i$ transforms only under the SU(2)$_L$ and the index $I$ under 
the SU(2)$_R$.
The super-Weyl transformations of $T_{iI}$ are clearly
\bea
(T_{iI})'=\re^{S}T_{iI}~.
\eea
To conclude note that, in terms of the chiral prepotential $W$ 
introduced in subsection \ref{subCTMII},
equations (\ref{descend-1}) and (\ref{descend-2}), 
the superfield $T_{iI}$ can be expressed in 
the following form 
\bea
T_{iI}&=&
{\ri\over 4}{[}\de_{+ i},\de_{- I}{]}W
={\ri\over 4}{[}\deb_{+ i},\deb_{- I}{]}\bar{W}
=(T^{iI})^*
~.
\label{T-W}
\eea

\subsection{2D $\cN=(4,4)$ covariant bi-projective superfields}
\label{cov2-proj}

In subsection \ref{TMII-2} we have rewritten the TM-II constraints in terms of analyticity like 
conditions on the left and right sectors of 2D $\cN=(4,4)$ supergravity.
Here we want to introduce a large class of analytic multiplets living in, what we call,
curved bi-projective superspace.

In defining curved bi-projective multiplets we follow the procedure recently developed in the 
cases of 5D $\cN=1$ supergravity \cite{KT-M_5D,KT-M_5Dconf} and 4D $\cN=2$ supergravity 
\cite{KLRT-M_4D-1,KLRT-M_4D-2}.
We then introduce isotwistors $u^{\opl}_i\in 
{\mathbb C}^2\setminus \{0\}$ and $v^{\bpl}_I\in {\mathbb C}^2\setminus\{0\}$ 
defined to be inert under the action of the structure group. 
In the present 2D $\cN=(4,4)$ case 
the difference compared with \cite{KT-M_5D,KT-M_5Dconf,KLRT-M_4D-1,KLRT-M_4D-2} 
is the use of two sets of 
isotwistor variables instead of one. This possibility is related to the fact that in
(\ref{TMII-analyticity}) we have two independent set of analyticity like constraints.
Note that the construction is based on and extends
the flat case of \cite{BusLinRoc,GHR,RSS,LR-biProj} 
and has clear similarities with the bi-harmonic superspace approach of 
\cite{IvanovSutulin,BellucciIvanov}.

Using the $u,v$ isotwistors we define the covariant derivatives
\bsubeq
\bea
&\de_+^\opl:=u^\opl_i\de_+^i~,~~~
\deb_+^\opl:=u^\opl_i\deb_+^i~,
\\
&\de_-^\bpl:=v^\bpl_I\de_-^I~,~~~
\deb_-^\bpl:=v^\bpl_I\deb_-^I~.
\eea
\esubeq
We are now ready to introduce a third equivalent definition of the covariant TM-II.
By contracting the $u,v$ isotwistors with $T_{iI}$ the superfield $T^{\opl\bpl}(z,u,v)$ is defined
according to the following equation
\bea
T^{\opl\bpl}(u,v):=u^\opl_i v^\bpl_IT^{iI}~.
\eea
The constraints (\ref{TMII-analyticity}) are then equivalent to
the analyticity like conditions
\bea
\de_+^\opl T^{\opl\bpl}=\deb_+^\opl T^{\opl\bpl}=0~,~~~~~~
\de_-^\bpl T^{\opl\bpl}=\deb_-^\bpl T^{\opl\bpl}=0~.
\eea
It is important to note that the superfield $T^{\opl\bpl}(u,v)$ is homogeneous of degree-(1,1) 
in the variables $u$ and $v$
\bea
T^{\opl\bpl}(c_Lu,v)=c_LT^{\opl\bpl}(u,v)~,~~
T^{\opl\bpl}(u,c_Rv)=c_RT^{\opl\bpl}(u,v)~,~~~
c_L,c_R\in{\mathbb C}\setminus\{0\}~.~~~~~~
\eea
In particular, $T^{\opl\bpl}$ describes an holomorphic tensor field on the product 
of two complex projective spaces ${\mathbb C}P^1\times{\mathbb C}P^1$.
The transformation rules of $T^{\opl\bpl}$ under $\bmL_{kl}, \bmR_{KL}$, that follow from 
(\ref{LTRT}),
can be written as
\bsubeq
\bea
\bmL_{kl}T^{\opl\bpl}(u^\opl,v^\bpl)&=&
-{1\over 2(u^{\opl}u^\omn)}\Big(u^{\opl}_{(k}u^{\opl}_{l)}D^{\omn\omn}
-u^{\opl}_{(k}u^\omn_{l)}\Big)
T^{\opl\bpl}(u^\opl,v^\bpl)~,
\label{LT}
\\
\bmR_{KL}T^{\opl\bpl}(u^\opl,v^\bpl)&=&
-{1\over 2(v^{\bpl}v^\bmn)}\Big(v^{\bpl}_{(K}v^{\bpl}_{L)}D^{\bmn\bmn}
-v^{\bpl}_{(K}v^\bmn_{L)}\Big)
T^{\opl\bpl}(u^\opl,v^\bpl)~,
\label{RT}
\eea
\esubeq
where we have introduced
\bsubeq
\bea
D^{\omn\omn}=u^{\omn i}{\pa\over \pa u^{\opl i}}~&,&~~~
D^{\bmn\bmn}=v^{\bmn I}{\pa\over \pa v^{\bpl I}}~,
\\
(u^{\opl}u^\omn):=u^{\opl i}u^\omn_i\ne 0~&,&~~~
(v^\bpl v^\bmn):=v^{\bpl I} v^\bmn_I \ne0~.
\label{uuvv}
\eea
\esubeq
The equations (\ref{LT}) and (\ref{RT}) involve two new isotwistors  $u^\omn_i$ and $v^\bmn_I$
which are subject to the only conditions (\ref{uuvv}) and are otherwise completely arbitrary.
The following relations also hold
\bea
C_{ij}={1\over (u^{\opl}u^\omn)}\big(u^{\opl}_ju^\omn_i-u^{\opl}_iu^\omn_j\big)~,~~~
C_{IJ}={1\over (v^{\bpl}v^\bmn)}\big(v^{\bpl}_Jv^\bmn_I-v^{\bpl}_Iv^\bmn_J\big)~.
\eea

The TM-II, in the form of $T^{\opl\bpl}$ just introduced, 
is the simplest example of a large class of multiplets living on 
$\cM^{2|4,4}\times {\mathbb C}P^1\times {\mathbb C}P^1$.
We call these bi-isotwistor superfields.\footnote{See
\cite{KT-M_5D,KT-M_5Dconf,KLRT-M_4D-1,KLRT-M_4D-2} for the introduction and examples of 
isotwistor superfields in 4D and 5D supergravities.}

A weight-(m,n) bi-isotwistor superfield $U^{(m,n)}(z,u^\opl,v^\bpl)$
 is holomorphic on an open domain of 
$\{{\mathbb C}^2 \setminus  \{0\}\}\times\{{\mathbb C}^2 \setminus  \{0\}\}$ 
 with respect to 
the homogeneous coordinates $(u^{\opl}_i,v^\bpl_I) $  
for ${\mathbb C}P^1\times {\mathbb C}P^1$,
and is characterized by the  conditions:\\
(i) it is  a homogeneous function of $(u^\opl,v^\bpl)$ 
of degree $(m,n)$, that is,  
\bsubeq
\bea
U^{(m,n)}(z,c_L\,u^\opl,v^\bpl)&=&(c_L)^m\,U^{(m,n)}(z,u^\opl,v^\bpl)~, 
\qquad c_L\in \mathbb{C}\setminus \{ 0 \}~,
\label{weight-m}
\\
U^{(m,n)}(z,u^\opl,c_R\,v^\bpl)&=&(c_R)^n\,U^{(m,n)}(z,u^\opl,v^\bpl)~, 
\qquad c_R\in \mathbb{C}\setminus \{ 0 \}~;
\label{weight-n}
\eea
\esubeq
(ii)  the supergravity gauge transformations act on $U^{(m,n)}$ 
as follows:
\bsubeq
\bea
\d_\cK U^{(m,n)} 
&=& \Big(K^{{C}} \de_{{C}} +K\cm+ (K_L)^{kl} {\bm L}_{kl}
+(K_R)^{KL}{\bm R}_{KL} \Big)U^{(m,n)} ~,  
\label{localU}
\\
{\bm L}_{kl}U^{(m,n)}&=& -\frac{1}{2(u^\opl u^\omn)} \Big(u^\opl_{(k}u^\opl_{l)} D^{\omn\omn} 
-m \, u^\opl_{(k}u^\omn_{l)} \Big) U^{(m,n)} ~,
\label{LU}
 \\ 
{\bm R}_{KL}U^{(m,n)}&=& -\frac{1}{2(v^\bpl v^\bmn)} \Big(v^\bpl_{(K}v^\bpl_{L)} D^{\bmn\bmn} 
-n \, v^\bpl_{(K}v^\bmn_{L)} \Big) U^{(m,n)} ~,
\label{RU}
\\
\cM U^{(m,n)}&=&{m-n\over 2} U^{(m,n)}~.
\label{LorentzU}
\eea 
\esubeq
Note that, due to (\ref{weight-m}),
the superfield $(\bmL_{kl}U^{(m,n)})$ is independent 
of $u^\omn_i$ even if the transformations in (\ref{LU}) explicitly depend on it; 
similarly $({\bm R}_{KL}U^{(m,n)})$ is independent of $v^\bmn_I$. 
We refer the reader to \cite{KLRT-M_4D-1} 
for a more detailed discussion on the 
SU(2) transformations of isotwistor-like superfields.

The most important property of 2D bi-isotwistor superfields is that the anticommutator among
any of the covariant derivatives 
$\de_+^\opl,\,\deb_+^\opl,\,\de_-^\bpl,\,\deb_-^\bpl$ is zero when acting on $U^{(m,n)}$.
Explicitly, it holds
\bea
0=\{\de_+^\opl,\de_+^\opl\}U^{(m,n)}
=\{\de_+^\opl,\deb_+^\opl\}U^{(m,n)}
=\{\de_+^\opl,\de_-^\bpl\}U^{(m,n)}
=\cdots~.~~~~~~
\label{dedeU}
\eea
We present a proof of this 
statement in appendix C.
It is worth mentioning that the Lorentz transformations of $U^{(m,n)}$ are uniquely fixed by 
requiring (\ref{dedeU}) with (\ref{LU}) and (\ref{RU}) assumed.
In the case in which $(m-n)$ is odd we will generically consider $U^{(m,n)}$ to be a 
fermionic superfield even if for the aim of the present discussion this is irrelevant.

With the definitions (i) and (ii) assumed, the set of bi-isotwistor superfields is closed under the 
product of superfields and the action of the $\de_+^\opl,\,\deb_+^\opl,\,\de_-^\bpl,\,\deb_-^\bpl$
derivatives.
More precisely, given a weight-(m,n) $U^{(m,n)}$ 
and a weight-(p,q)  $U^{(p,q)}$ bi-isotwistor superfields 
 the superfield $(U^{(m,n)}U^{(p,q)})$ is a 
weight-(m+p,n+q) bi-isotwistor superfield.
Moreover, the superfields $(\de_+^\opl U^{(m,n)})$, $(\deb_+^\opl U^{(m,n)})$ and 
$(\de_-^\bpl U^{(m,n)})$, $(\deb_-^\bpl U^{(m,n)})$ are respectively weight-(m+1,n) and 
weight-(m,n+1) bi-isotwistor superfields.

If we consider the set of bi-isotwistor superfields transforming homogeneously 
under super-Weyl transformations $(U^{(m,n)})'=\re^{w S}U^{(m,n)}$, it is natural to impose
\bea
 (U^{(m,n)})'=\re^{\frac{m+n}{2}S}U^{(m,n)}~.
\label{USW}
\eea
The conformal weight $w$ in the previous relation is fixed by the requirement that 
the superfields 
$\de_+^\opl U^{(m,n)},\,\deb_+^\opl U^{(m,n)}\,,\de_-^\bpl U^{(m,n)},\,\deb_-^\bpl U^{(m,n)}$
also transform homogeneously. 
For example, it holds
\bea
(\de_+^\opl U^{(m,n)})'&=&
\re^{{m+n+1\over 2} S}\de_+^\opl U^{(m,n)}~.
\eea
To prove the last relation one needs to use the equations (\ref{SW+}), (\ref{USW}),
(\ref{LorentzU})
and the relation
\bea
u^{\opl k}\bmL_{kl}U^{(m,n)}={m\over 2}u^\opl_lU^{(m,n)}~,
\eea
which follows from eq. (\ref{LU}).
Analogously one can prove that also the superfields 
$\deb_+^\opl U^{(m,n)},\,\de_-^\bpl U^{(m,n)}$ and $\deb_-^\bpl U^{(m,n)}$ 
have conformal weight $w=(m+n+1)/2$ if eq. (\ref{USW}) is assumed.

We are now ready to introduce 2D $\cN=(4,4)$ covariant bi-projective superfields.
We define a weight-(m,n) covariant bi-projective supermultiplet 
$Q^{(m,n)}(z,u^\opl,v^\bpl)$ to be a bi-isotwistor superfield satisfying (i), (ii), 
(\ref{weight-m})--(\ref{LorentzU})
and to be constrained by the analyticity  conditions
\be
\de^\opl_{+} Q^{(m,n)}  =\deb^\opl_{+} Q^{(m,n)}=0~,~~~~~~
\de^\bpl_{-} Q^{(m,n)}  =\deb^\bpl_{-} Q^{(m,n)}=0~.
\label{analyticity}
\ee  
Note that the consistency of the previous constraints is guaranteed by eq. (\ref{dedeU}).
This now takes the form of an integrability condition for the 
analyticity constraints.

If we ask $Q^{(m,n)}$ to have homogeneous super-Weyl transformations, it is clear 
by the previous discussion on the super-Weyl transformations of bi-isotwistor superfields,
that the transformations
\bea
 (Q^{(m,n)})'=\re^{\frac{m+n}{2}S}Q^{(m,n)}~,
\label{QSW}
\eea
preserve the analyticity conditions (\ref{analyticity}).

Given a bi-projective multiplet $Q^{(m,n)}(z,u^\opl,v^\bpl)$,
its complex conjugate 
is not covariantly analytic.
However, one can introduce a generalized,  analyticity-preserving 
conjugation, $Q^{(m,n)} \to \widetilde{Q}^{(m,n)}$, defined as
\bsubeq
\bea
&&\widetilde{Q}^{(m,n)} (u^\opl,v^\bpl)\equiv \bar{Q}^{(m,n)}\big(
\overline{u^\opl}\to 
\widetilde{u}^\opl,
\overline{v^\bpl}\to 
\widetilde{v}^\bpl\big)~, 
\\
&&
\widetilde{u}^\opl = {\rm i}\, \s_2\, u^\opl~, 
~~~
\widetilde{v}^\bpl = {\rm i}\, \s_2\, v^\bpl~, 
\eea
\esubeq
with $\bar{Q}^{(m,n)}(\overline{u^\opl},\overline{v^\bpl})$ the complex conjugate of $Q^{(m,n)}$
and $\overline{u^\opl},\overline{v^\bpl}$ the complex conjugates of $u^\opl,v^\bpl$.
It is easy to check that $\widetilde{Q}^{ (m,n) } (z,u^\opl,v^\bpl)$ 
is a  weight-(m,n) bi-projective multiplet.
One can see that
$\widetilde{\widetilde{Q}}{}^{(m,n)}=(-1)^{m+n}Q^{(m,n)}$,
and therefore real supermultiplets can be consistently defined when 
$(m+n)$ is even.
The superfield $\widetilde{Q}^{(m,n)}$ is called the smile-conjugate of 
${Q}^{(m,n)}$. Geometrically, this conjugation is complex conjugation composed
with the antipodal map on the two projective spaces ${\mathbb C}P^1\times {\mathbb C}P^1$.
The simplest example of real bi-projective superfield is again the TM-II.
The reality condition $(T_{iI})^*=T^{iI}$ is equivalent to 
$\widetilde{T}^{\opl\bpl}=T^{\opl\bpl}$.

Note that,  by definition, the TM-II superfield $T^{\opl\bpl}$ describes a regular holomorphic tensor 
field on the whole product of the two complex projective spaces 
$\mathbb{C}P^1\times\mathbb{C}P^1$.
Other simple examples of bi-projective superfields that are regular holomorphic tensor field
on the whole $\mathbb{C}P^1\times\mathbb{C}P^1$ can be given by what we call
$O(m,n)$ multiplets ($m,n>0$). They are described by a bi-projective superfield
$O^{(m,n)}(z,u,v):=u^\opl_{i_1}\cdots u^\opl_{i_m}v^\bpl_{J_1}\cdots v^\bpl_{J_n}
O^{i_1\cdots i_m J_1\cdots J_n}(z)$, where the isotensor superfield
$O^{i_1\cdots i_m J_1\cdots J_n}={1\over m! n!}O^{(i_1\cdots i_m) (J_1\cdots J_n)}$ 
is such that 
\bsubeq
\bea
\cm O^{i_1\cdots i_m J_1\cdots J_n}
&=&{m-n\over 2}O^{i_1\cdots i_m J_1\cdots J_n}
~,
\label{M-O}
\\
\bmL_{kl}O^{i_1\cdots i_m J_1\cdots J_n}
&=&-\hf{1\over (m-1)!}\d^{(i_1}_{(k}O_{l)}{}^{i_2\cdots i_m) J_1\cdots J_n}
~,
\label{L-O}
\\
\bmR_{KL}O^{i_1\cdots i_m J_1\cdots J_n}
&=&-\hf{1\over (n-1)!}\d^{(J_1}_{(K}O^{|i_1\cdots i_m|}{}_{L)} {}^{J_2\cdots J_n)}
~,
\label{R-O}
\\
\de_+^{(k}O^{i_1\cdots i_m) J_1\cdots J_n}&=&
\deb_+^{(k}O^{i_1\cdots i_m) J_1\cdots J_n}=0
~,
\label{an1-O}
\\
\de_-^{(K}O^{|i_1\cdots i_m| J_1\cdots J_n)}&=&
\deb_-^{(K}O^{|i_1\cdots i_m| J_1\cdots J_n)}=0
~.
\label{an2-O}
\eea
\esubeq
Note that $O^{(m,n)}(u,v)$ is polynomial in the isotwistor variables $u,\,v$.
More general bi-projective multiplets have poles and more complicate analytic properties 
on $\mathbb{C}P^1\times\mathbb{C}P^1$. 
Then, in general 2D covariant bi-projective superfields possess an infinite 
number of standard superfields in a way completely analogue to the more studied 4D-5D 
curved cases \cite{KT-M_5D,KT-M_5Dconf,KLRT-M_4D-1,KLRT-M_4D-2}.
A more detailed classification of covariant bi-projective superfields will be considered elsewhere.

One can represent a bi-projective superfield $Q^{(m,n)}$
in the form
\bea
Q^{(m,n)}=
-{1\over 4}\de_{+}^\opl\de_{-}^\bpl\deb_{+}^\opl\deb_{-}^\bpl \cU^{(m-2,n-2)}
={1\over 4}\de_{+}^\opl\de_{-}^\bpl\deb_{-}^\bpl \deb_{+}^\opl \cU^{(m-2,n-2)}
=\cdots~,
\label{Projector-Q}
\eea
for some bi-isotwistor superfield $\cU^{(m-2,n-2)}$ satisfying
(\ref{weight-m})--(\ref{LorentzU}) and (\ref{USW}).
In (\ref{Projector-Q}), thanks to (\ref{dedeU}) and the defining properties of bi-isotwistor 
superfields, one can take any graded permutation of the covariant derivatives
$\de_+^\opl,\,\deb_+^\opl,\,\de_-^\bpl,\,\deb_-^\bpl$.
Therefore it is trivial to prove that (\ref{analyticity}) is identically satisfied.
We will call a $\cU^{(m-2,n-2)}$, such that (\ref{Projector-Q}) holds, a bi-isotwistor prepotential 
of $Q^{(m,n)}$.

To conclude we observe that all the results presented in this subsection remain true, up to
few minor differences, if one reduces the supergravity geometry to the minimal multiplet.
Due to the de-gauging from the SU(2)$_L\times$SU(2)$_R$ group
to SU(2)$_\cV$,
 in the minimal case,
the supergravity gauge transformations of bi-isotwistor and bi-projective superfields
are modified from eq. (\ref{localU}) to
$\d_\cK U^{(m,n)} = \Big(K^{{C}} \de_{{C}} +K\cm+ (K_\cV)^{kl} {\cV}_{kl}\Big)U^{(m,n)}$.
Note that eqs. (\ref{LU})--(\ref{LorentzU}) remain the same.
A second modification that occurs 
regards the super-Weyl transformations.
As explained in subsection 2.3, to preserve the gauge (\ref{minimalGAUGE}),
in the minimal case the super-Weyl transformations are generated by a TM-II $(\bS,\bS_{ij})$
couple of superfields through $\tilde{\d}=(\d-\d_\cC)$ infinitesimal transformations.
For bi-isotwistor and bi-projective superfields this means that
the infinitesimal super-Weyl transformations are modified to
$ \tilde{\d}U^{(m,n)}=\big(\frac{m+n}{2}\bS  -\bS^{kl}\cC_{kl}\big)U^{(m,n)}$
that includes a compensating SU(2)$_\cC$ transformation.

\subsection{Action principle}

Here we give a bi-projective superfield action principle invariant under the supergravity gauge 
group and super-Weyl transformations and such that in the flat limit it reduces to the one 
introduced in \cite{BusLinRoc,LR-biProj}.

Let $\cL^{(0,0)}$ be a real bi-projective superfield of weight-$(0,0)$.
In particular, according to (\ref{USW}), $\cL^{(0,0)}$ is invariant under super-Weyl transformations. 
Moreover, we consider a TM-II described by $T^{\opl\bpl}$ with $W, (\bar{W})$ a chiral  
prepotential.
Associated with $\cL^{(0,0)}$ we introduce the action principle
\bea
S&=&{1\over 4 \pi^2}
 \oint (u^\opl \rd u^\opl)
  \oint (v^\bpl \rd v^\bpl)
\int \rd^2 x \,{\rm d}^8\q \,E\, \frac{W{\bar W}}{(T^{\opl\bpl})^2}\cL^{(0,0)}~, 
~~~ E^{-1}= {\rm Ber}(E_A{}^M)~.
~~~~~~
\label{InvarAc}
\eea
By construction, the functional is  invariant under the  re-scaling
$u_i^\opl(t)  \to c_L(t) \,u^\opl_i(t) $, for an arbitrary function
$ c_L(t) \in {\mathbb C}\setminus  \{0\}$, 
where $t$ denotes the evolution parameter 
along the first closed integration contour.\footnote{For simplicity in this paper we consider 
the two contour integrals to be closed. Depending on the explicit form of the Lagrangian 
$\cL^{(0,0)}$, one could consider different cases \cite{GHR,BusLinRoc,RSS,LR-biProj}
with line integrals not necessarily closed.}
 Similarly, (\ref{InvarAc}) is invariant
under  re-scalings
$v_I^\bpl(s)  \to c_R(s) \,v^\bpl_I(s) $, for an arbitrary function
$ c_R(s) \in {\mathbb C}\setminus  \{0\}$, 
where $s$ denotes the evolution parameter 
along the second closed integration contour.
Note that (\ref{InvarAc}) has clear similarities with the action principles in four and five-dimensional
curved projective superspace  \cite{KT-M_5D,KT-M_5Dconf,KLRT-M_4D-1,KLRT-M_4D-2}.

By using that under super-Weyl transformations  $E$ transforms like
\bea
E'=\re^{2S}E~,
\eea
and the transformations $(T^{\opl\bpl})'=\re^{S}T^{\opl\bpl}$ and $W'=W$, 
one sees that $S$ is super-Weyl invariant. 
The action (\ref{InvarAc}) is also invariant under arbitrary local supergravity gauge 
transformations (\ref{SUGRAgauge1})--(\ref{tensor-K}). 
The invariance under general coordinates and Lorentz transformations is trivial.
The invariance under the two SU(2) transformations can be proved similarly to \cite{KT-M_5D}.
It is instructive to review this in the 2D case.

The proof of SU(2) invariance goes as follows. Under infinitesimal SU(2)$_L$ transformations the 
action varies like
\bea
\d_LS&=&
{1\over 4\pi^2}
 \oint (u^\opl \rd u^\opl)
  \oint (v^\bpl \rd v^\bpl)
\int \rd^2 x \,{\rm d}^8\q \,E\, W{\bar W}
(K_L)^{kl}\bmL_{kl}\Big(\frac{\cL^{(0,0)}}{(T^{\opl\bpl})^2}\Big)~,
\label{Lvar}
\eea
where we have used the invariance of $E$ and $W$ under SU(2) transformations.
For weight-(-2,-2) bi-projective superfields, like $Q^{(-2,-2)}:= (\cL^{(0,0)})/{(T^{\opl\bpl})^2}$, 
 it holds
\bea
(K_L)^{kl} \bmL_{kl} \, Q^{(-2,-2)}= -\frac{1}{(u^\opl u^\omn)} D^{\omn\omn}
\Big( (K_L)^{\opl\opl} Q^{(-2,-2)}\Big)~.
\eea
Next, note that the $(u^\opl \rd u^\opl)$ integration measure, written in terms of the evolution 
parameter $t$ of the closed contour, is equal to
\bea
(u^\opl\rd u^\opl)=-(\dt{u}^\opl u^\opl)\rd t~,~~~\dt{f}:=\rd f(t)/\rd t~.
\eea
Then, being $(K_L)^{\opl\opl} Q^{(-2,-2)}$ homogeneous of degree zero in $u^\opl_i$ 
it is easy to note that it holds
\bea
(u^\opl \rd u^{\opl} )\, (K_L)^{kl} \bmL_{kl}  \,Q^{(-2,-2)}= -{\rm d}t \,
\frac{{ \rm d}  }{{\rm d}t} \, \Big((K_L)^{\opl\opl} Q^{(-2,-2)}\Big)~.
\eea
Since the integration contour is closed, eq. (\ref{Lvar}) is zero and the SU(2)$_L$-part of 
the supergravity transformations does not contribute to the variation of 
the action (\ref{InvarAc}). 
The proof of the invariance under the SU(2)$_R$ transformations goes along the same lines.

If, according to eq. (\ref{Projector-Q}),
we represent $\cL^{(0,0)}$ in terms of a bi-isotwistor prepotential $\cU^{(-2,-2)}$,
then the action (\ref{InvarAc}) can be rewritten as
\bea
S&=&
{1\over 4\pi^2}
 \oint (u^\opl \rd u^\opl)
  \oint (v^\bpl \rd v^\bpl)
\int \rd^2 x \,{\rm d}^8\q \,E\, \cU^{(-2,-2)}~.
\label{InvarAc2}
\eea
Here we have used the relations
$T^{\opl\bpl}=({\ri/4})[\de_+^\opl,\de_-^\bpl]W=({\ri/4})[\deb_+^\opl,\deb_-^\bpl]\bar{W}$
that follow from (\ref{T-W}), and 
\bea
\deb_-^\bpl \deb_+^\opl\de_-^\bpl  \de_+^\opl (W\bar{W})=-4 (T^{\opl\bpl})^2~.
\label{useful3}
\eea
After integrating by parts, one obtains eq. (\ref{InvarAc2}) from (\ref{InvarAc}).
The equation (\ref{InvarAc2}) leads to an important result: if $\cL^{(0,0)}$ 
is a function of some 
supermultiplets to which the TM-II compensator does not belong, then the 
action $S$ is independent of the superfields  $T^{\opl\bpl}$, $W$ and $\bar{W}$ chosen.

Let us take the flat limit of the action principle (\ref{InvarAc}). This is 
\bea
S_{\rm flat}&=&
{1\over 4\pi^2}
 \oint {(u^\opl \rd u^\opl)\over (u^\opl u^\omn)^2}
  \oint {(v^\bpl \rd v^\bpl)\over (v^\bpl v^\bmn)^2}
\int \rd^2 x \,D_+^\omn \DB_+^\omn D_-^\bmn \DB_-^\bmn
D_+^\opl \DB_+^\opl D_-^\bpl \DB_-^\bpl
   \frac{W{\bar W}}{(T^{\opl\bpl})^2}L^{(0,0)}~,
   ~~~~~~~
\eea
where $D_{\a i},\DB_\a^i$ are the flat covariant derivatives, $L^{(0,0)}$ is the lagrangian in the flat 
case and 
$D^\omn_+=u_i^\omn D_+^i$, $\DB^{\omn}_+=u_i^\omn\DB^i_+$,
$D^\bmn_-=v_I^\bmn D_-^I$ and $\DB^{\bmn}_-=v_I^\bmn\DB^I_-$.
Using analyticity of $T^{\opl\bpl}$ and of the Lagrangian $L^{(0,0)}$, and
the relation (\ref{useful3}) in the flat limit, we obtain
\bea
S_{\rm flat}&=&
{1\over \pi^2}
 \oint {(u^\opl \rd u^\opl)\over (u^\opl u^\omn)^2}
  \oint {(v^\bpl \rd v^\bpl)\over (v^\bpl v^\bmn)^2}
\int \rd^2 x \,D_+^\omn \DB_+^\omn D_-^\bmn \DB_-^\bmn
L^{(0,0)}~.
~~~~~~
\label{Sflat}
\eea
In the north chart of both the ${\mathbb C}P^1$ one can obtain the flat action 
principle written in terms of inhomogeneous coordinates for 
${\mathbb C}P^1\times{\mathbb C}P^1$. 
This coincides with the action principle given in \cite{BusLinRoc,LR-biProj}. 
The action (\ref{Sflat}) is a 2D 
analogue of the 4D case of \cite{KLR,LR} and, being written in homogeneous coordinates 
for the projective spaces, it is closer in form to the one in \cite{Siegel,Siegel2DN4}.
The action (\ref{Sflat}) is also invariant 
under arbitrary ``projective'' transformations of the form:
\begin{subequations}
\bea
(u_i{}^\omn\,,\,u_i{}^\opl)~\to~(u_i{}^\omn\,,\, u_i{}^\opl )\,P_L~,~~~~~~P_L\,=\,
\left(\begin{array}{cc}a_L~&0\\ b_L~&c_L~\end{array}\right)\,\in\,{\rm GL(2,\mathbb{C})}~,
\label{projectiveGaugeVar-L}
\\
(v_I{}^\bmn\,,\,v_I{}^\bpl)~\to~(v_I{}^\bmn\,,\, v_I{}^\bpl )\,P_R~,~~~~~~P_R\,=\,
\left(\begin{array}{cc}a_R~&0\\ b_R~&c_R~\end{array}\right)\,\in\,{\rm GL(2,\mathbb{C})}~.
\label{projectiveGaugeVar-R}
\eea
\end{subequations}
Projective transformations express the homogeneity of the formalism with respect to
 $u^\opl,\,v^\bpl$ 
and the independence on $u^\omn,\,v^\bmn$. This invariance results a powerful tool in 
superspace theories with eight supercharges. For example, in 5D $\cN=1$ \cite{KT-M_5D}
and 4D $\cN=2$ \cite{KT-M-normal} supergravity it has been used to reduce the projective action 
principle to components.
Along the same lines, one could approach the 2D $\cN=(4,4)$ case
and continue the analysis of component reduction of 2D $\cN=(4,4)$
superspace action principles of \cite{GatesGT-M-09-1,GatesMorrison}.

We conclude by noting that the action (\ref{InvarAc}) has the same form 
if one considers the
supergravity geometry reduced to the minimal multiplet. 
In such case (\ref{InvarAc})
is invariant under arbitrary supergravity gauge transformations, with
SO(1,1)$\times$SU(2)$_\cV$ as tangent space group. It is clearly invariant also under
the $\tilde{\d}$ variation of subsection 2.3 generated by $\bS,\,\bS_{ij}$ which includes
super-Weyl and compensating SU(2)$_\cC$ transformations.

We believe that the action (\ref{InvarAc}) is suitable to describe general 2D $\cN=(4,4)$
superconformal matter systems, such as  WZNW, Liouville systems and
 non-linear sigma models, covariantly coupled to supergravity. 
The investigation of that subjects and a more detailed study of bi-projective multiplets
is left for future research.


\section{Conclusion}
\setcounter{equation}{0}

In this paper we presented new results in the study of 2D $\cN=(4,4)$ supergravity
using superspace techniques.
We proposed a new superspace formulation for $\cN=(4,4)$ conformal supergravity in 
two dimensions
which proves 
to be an extension of the minimal multiplet of \cite{2DN4SG}.
We then
described the covariant coupling of supergravity to a large class of multiplets.
We begun by coupling the extended supergravity to an Abelian vector multiplet
 described by a twisted-II multiplet. 
We have then introduced 
so called covariant bi-projective supermultiplets 
and presented a manifestly locally supersymmetric and super-Weyl 
invariant action principle in bi-projective superspace.

The formalism we have introduced
should be suitable to study general classes
of matter couplings in 2D $\cN=(4,4)$ supergravity.
In the superspace supergravity framework presented here, 
possible subjects for future investigations would be the formulation of 
2D $\cN = (4, 4)$ super-conformal matter systems such as WZNW/Liouville-type systems,
non-linear sigma models and $(4,4)$ non-critical strings.

We also believe that there are still open questions purely related to 
2D $\cN = (4, 4)$ supergravity in superspace.
One first question is the existence of variant minimal formulations.
The multiplet we presented in the paper is an extension of the minimal multiplet of \cite{2DN4SG}
but our analysis indicate that the latter is the only minimal de-gauging 
of the supergravity of subsection \ref{ExtendedSUGRA}.
It is natural to believe that there exists 
another minimal formulation having TM-II torsion components and
TM-I conformal compensator. 
Such new minimal multiplet would turn out to be dual to the one of \cite{2DN4SG}
in a manner similar to the $\cN=(2,2)$ case of 
\cite{HowePapa1987,2DN4SG,Grisaru:1994dm,Gates:1995du}.
One way to find it could be by dimensional reduction of 4D $\cN=2$ superspace supergravity.
An alternative approach could be the study of a non-minimal supergravity
in which the structure group of the curved superspace
is  the full automorphism group of 
$\cN=(4,4)$ supersymmetry \cite{Siegel2DN4}:
SO(1,1)$\times$SO(4)$_L\times$SO(4)$_R$.
In this paper, 
we didn't 
considered the extra 
SU(2)$_L\times$SU(2)$_R$, which, for example, transform $\de_{+i}$ into $\deb_{+ i}$.
It would be useful to rewrite our results in a basis of derivatives 
$(\de_{+ \underline{i}i},\,\de_{-\underline{I}I})$ where the SO(4)$_L\times$SO(4)$_R$ structure is 
manifest; the new minimal multiplet could probably be identical to the one of \cite{2DN4SG}
but with simply the non-underlined groups and indices changed with the underlined ones. 
This covariant derivative 
basis would help to compare in details our discussion with the bi-harmonic superspace
results of \cite{IvanovSutulin,BellucciIvanov}.

Clearly the solution of the constraints of the 2D $\cN=(4,4)$ supergravity multiplets
would be interesting.
Note that a first, but uncompleted, effort to solve the minimal constraints 
in terms of prepotentials was given in \cite{KetovPrep2DN4} along the lines of
the 2D $\cN=(2,2)$ case of \cite{Grisaru:1994dm}.
 A formulation of  2D $\cN=(4,4)$ conformal 
supergravity purely based on prepotentials
lies in the bi-harmonic superspace approach of \cite{BellucciIvanov}.
A complete solution of the  Wess-Zumino like constraints for the minimal, or non-minimal,
multiplet could clarify the connection between 
the bi-projective and bi-harmonic superspace approaches
and provide an understanding of 2D $\cN=(4,4)$
supergravity in the spirit of the Gates-Siegel prepotential approach to the 4D
$\cN=1$ case \cite{SG}.
To this regard the 2D case would be of example for the higher dimensional cases where the 
structure of the supergravity multiplets is more involved.

The reduction to subsuperspaces is another topic in which the 2D $\cN=(4,4)$ case could be 
fruitful to clarify more involved higher-dimensional cases.
A detailed analysis of bi-projective superfields and the action principle reduced to 2D 
$\cN=(2,2)$ superspace would be very interesting.
%
~~~\\

\noindent
{\bf Acknowledgements:}\\
We are grateful to S. James Gates, Jr. to have raised our interest on 2D $\cN=(4,4)$ superspace 
supergravity and for many useful comments and discussions.
We also thank S. M. Kuzenko for comments.
We are grateful to the School of Physics at the University of Western Australia for hospitality and 
support during August 2009.
We thank the organizers of the Workshop ``Supersymmetries and Quantum Symmetries''
(SQSÕ09), at the Bogoliubov Laboratory of Theoretical Physics, JINR, Dubna, 
July 29--August 3 2009,
for hospitality.
The author is also grateful to Konstantinos Koutrolikos and Simon Tyler
for reading the manuscript.

This research was supported  by the endowment 
 of the John S.~Toll Professorship, the University of 
 Maryland Center for String \& Particle Theory, and
 National Science Foundation Grant PHY-0354401.


\appendix
\section{2D Conventions}
\setcounter{equation}{0}

In this section we collect the two dimensional conventions used in the paper.
These are consistent with \cite{SUPERSPACE,GatesMorrison}.
The Minkowski metric, Levi-Civita tensor and $\g$-matrices in two dimensions are defined 
according to the equations
\begin{subequations}
\bea
& \eta_{ a b} = (1 , -1)  ~,~~~ \ve_{a b} \ve^{c d} ~=~  
- \d_{[a}^c  \d_{b]}^d  ~,~~~ \ve^{0 1} =  +1 ~, 
  \\ &
  (\g^a)_{\a}{}^{ \g} (\g^b)_{\g}{}^{ \b} =   \eta^{ a b} \d_{\a}^{\b}
- \ve^{ a b}  (\g^3)_{\a}{}^{\b} ~, 
\label{gamma-1}
\eea
\end{subequations}
where the Lorentz spinor indices take values $\a=+,-$.
It is important to remark that in this paper, the complete (anti)symmetrization
of $n$ indices does not involve any $(1/n!)$ factor.\footnote{For example, our conventions tell that 
$\psi_{(\a}\chi_{\b)}=(\psi_{\a}\chi_{\b}+\psi_{\b}\chi_{\a})$ 
and $\psi_{[\a}\chi_{\b]}=(\psi_{\a}\chi_{\b}-\psi_{\b}\chi_{\a})$.}
Equation (\ref{gamma-1}) imply
\bsubeq
\bea
&(\g^a)_\a{}^\g (\g_{a})_\g{}^\b = 2 \d_\a^\b ~, ~~~
(\g^3)_\a{}^\g (\g^{a})_\g{}^\b =- \ve^{a b} (\g_{b})_\a{}^\b ~,
\\
&(\g^3)_\a{}^\b(\g^a)_\b{}^\a=0~,
~~~
(\g^3)_\a{}^\g(\g^3)_\g{}^\b=
\d_\a^\b~.
\eea
\esubeq

Some Fierz identities used in the paper are:
\begin{subequations}
\bea
& C_{\a \b}  C^{\g \d} =~ \d_{[\a}^\g  \d_{\b]}^\d ~,
\\
&(\g^a)_{\a \b} (\g_a)^{\g \d} + (\g^3)_{\a \b} (\g^3)^{\g \d}
= - \d_{(\a}^\g  \d_{\b)}^\d  ~,
\\ &
 (\g^a)_{(\a}{}^\g (\g_a)_{\b)}{}^\d + (\g^3)_{(\a}{}^\g 
(\g^3)_{\b)}{}^\d = \d_{(\a}^\g  \d_{\b)}^\d    ~, 
\\ &
(\g^a)_{(\a}{}^\g (\g_a)_{\b)}{}^\d = -2 (\g^3)_{\a \b} 
(\g^3)^{\g \d} ~, 
\\ &
2 (\g^a)_{\a \b} (\g_a)^{\g \d} + (\g^3)_{(\a}{}^\g (\g^3)_{\b)}{}^\d   = - \d_{(\a}^\g  \d_{\b)}^\d ~,
\\ &
 (\g_a)_\a {}^\d \d_\b {}^\g + (\g^3 \g_a)_\a {}^\g (\g^3)_\b {}^\d
= (\g^3 \g_a)_{\a \b} (\g^3)^{\g \d} ~, 
\\ &
  (\g^3 \g_a\g^3)_\a{}^{\d}= -(\g_a)_\a {}^\d~,   
\\
&(\g^c)_\a{}^\rho(\g^3\g_c)_{\d}{}^{\b}=
C_{\a\d}(\g^3)^{\b \rho} 
 + (\g^3)_{\a\d}  C^{\rho\b} 
 ~.
 \eea
\end{subequations}
In some of the previous relations, given for example a spinor $\psi^\a(x)$,
we have raised and lowered the spinor indices according to the rule
\bea
 \psi^{\a}(x) = C{}^{\a\b} \,  \psi_{\b}(x)  ~,~~~
  \psi_{\a}(x) =  \psi^{\b}(x)  \,  C{}_{\b\a}  ~.
  \eea

In terms of an explicit representation, we can define the 2D $\g$-matrices
by using the usual Pauli matrices according to
\bea
(\g^0 )_{\a}{}^{\b} \equiv (\s^2  )_{\a}{}^{\b}  ~,~~~ 
(\g^1)_{\a}{}^{\b} \equiv - i (\s^1  )_{\a}{}^{\b} ~,~~~
(\g^3)_\a{}^\b \equiv (\s^3  )_{\a}{}^{\b} ~~~.  
\eea
The spinor metric
$C_{\a \b}$ and its inverse $C^{\a \b}$ can be defined by
\bea
C_{\a \b} \equiv (\s^2 )_{\a \b} ~,~~~ 
C^{\a \b} \equiv- (\s^2 )^{\a \b} ~.
\eea
Using this explicit representation, it is easy to show the following
symmetry properties
\bea
(\g^a)_{\a \b} ~=~ (\g^a)_{\b \a} ~~~,~~~
(\g^3)_{\a \b} ~=~ (\g^3)_{\b \a} ~~~, ~~~
C_{\a \b} ~ ~=~ - C_{\b \a} ~, 
\eea
and similarly for the matrices with both up indices.
The following complex conjugation properties can be derived
\begin{subequations}
\bea
&((\g^a)_{\a}{}^{ \b} )^* = -  (\g^a)_{\a}{}^{ \b} ~,~~~
((\g^3)_{\a}{}^{ \b} )^* =   (\g^3)_{\a}{}^{ \b} ~,~~~
( C_{\a \b}  )^* ~=~ - C_{\a \b} ~,
\\
&( (\g^a)_{\a \b}  )^* = (\g^a)_{\a \b} ~,~~~
( (\g^3)_{\a \b}  )^* = -(\g^3)_{\a \b} ~,
\eea
\end{subequations}
and the same for the matrices with both indices raised.
The choice of gamma matrices is
in a Majorana representation and the simplest spinor one can choose is real $\psi^{\a}(x)$, 
\bea
(\psi^{\a}(x) )^* = \psi^{\a}(x) ~,~~~
(\psi_{\a}(x))^* =   -  \psi_{\a}(x)  ~.
 \eea
Clearly it is also possible to introduce complex spinors.

The SU(2) indices $i=\1,\,\2$ possess conventions similar to the one
used for the Lorentz spinor indices.  The SU(2) metric
$C_{i j}$ and its inverse $C^{ij}$ satisfy
\bsubeq
\bea
&C_{i  j} \equiv (\s^2 )_{i  j} ~,~~~ 
C^{i  j} \equiv- (\s^2 )^{i  j} ~,  
\\
&C_{i j} =  - C_{j i} ~,~~~
C^{ij} =  -  C^{j i} ~,~~~
C_{i  j}  C^{k l} = \d_{[i}^{k} \d_{j]}^{l}~.
\eea
\esubeq
We raise and lower SU(2)  indices according to
\bea
 \psi^{i}(x) ~=~ C{}^{i  j} \,  \psi_{j}(x)  ~~~,~~~
  \psi_{i}(x) ~=~  \psi^{j}(x)  \,  C{}_{j i}  ~.
\eea
Note that for the SU(2) invariant it holds
\bea
(C^{ij})^*=C_{ij}~,~~~
(C_{ij})^*=C^{ij}~.
\eea
With the previous complex conjugation conventions we have that the local Grassmanian 
superspace coordinates $(\q^{\mu \imath}, \qb^\mu_\imath)$ are related one to each other by
the rule 
\bea
(\q^{\mu \imath})^*=\qb^\mu_{\imath}~.
\eea
Accordingly, given a general complex superfield $A$ ($\bar{A}:=(A)^*$), 
with Grassmann parity $\ve(A)$,
the complex conjugate of the spinor covariant derivatives of $A$ satisfies
\bea
(\de_{\a i}A)^*=-(-)^{\ve(A)}\deb_\a^i\bar{A}~.
\eea

To conclude, let us 
give the commutation algebra of the SO(1,1)$\times$SU(2)$_L\times$SU(2)$_R$
generators $\cM$, $\bmL_{kl}$ and $\bmR_{kl}$
which can be derived by using (\ref{Lorentz1})--(\ref{R_kl})\bsubeq\bea
&&{[}\cM,\cM{]}={[}\cM,\bmL_{kl}{]}={[}\cM,\bmR_{kl}{]}={[}\bmL_{ij},\bmR_{kl}{]}=0~,
\label{[MMLR]}
\\
&&~~~~~~~~~\,
{[}\bmL_{ij},\bmL_{kl}{]}=
{1\over 4}\big(
C_{k(i}\bmL_{j)l}
+C_{l(i}\bmL_{j)k}
\big)
~,
\label{[LL]}
\\
&&~~~~~~~~~
{[}\bmR_{ij},\bmR_{kl}{]}=
{1\over 4}\big(
C_{k(i}\bmR_{j)l}
+C_{l(i}\bmR_{j)k}
\big)
~.
\label{[RR]}
\eea
\esubeq
The commutation algebra for the operators $\cV_{kl}=(\bmL_{kl}+\bmR_{kl})$ and 
$\cC_{kl}=(\bmL_{kl}-\bmR_{kl})$ is
\bsubeq
\bea
{[}\cV_{ij},\cV_{kl}{]}&=&
{1\over 4}\Big(
C_{k(i}\cV_{j)l}
+C_{l(i}\cV_{j)k}
\Big)
~,
\label{[VV]}
\\
{[}\cV_{ij},\cC_{kl}{]}&=&
{1\over 4}\Big(
C_{k(i}\cC_{j)l}
+C_{l(i}\cC_{j)k}
\Big)
~,
\label{[VC]}
\\
{[}\cC_{ij},\cC_{kl}{]}&=&
{1\over 4}\Big(
C_{k(i}\cV_{j)l}
+C_{l(i}\cV_{j)k}
\Big)
~.
\label{[CC]}
\eea
\esubeq


\section{Solution of the supergravity Bianchi identities}
\label{SUGRA-Bianchi}
\setcounter{equation}{0}

In this appendix we want to give a description of the solution of the Bianchi identities for the 
2D $\cN=(4,4)$ supergravity of subsection \ref{ExtendedSUGRA}
based on the torsion constraints (\ref{constr-0})--(\ref{constr-1}).
In a standard and useful way the analysis is organized in accordance with the increasing mass 
dimension of the Bianchi identities involved.

The super-Jacobi identities for the covariant derivatives
\bea
\sum_{[ABC)}{[}\de_{{A}},{[}\de_{{B}},\de_{{C}}\}\}
~=~0~,
\eea
with the graded cyclic sum assumed, are equivalent to the following Bianchi identities
for the torsion and curvature of the geometry
\bsubeq
\bea
0&=&\sum_{[ABC)}\Big(R_{AB}{}_C{}^D-\de_AT_{BC}{}^D+T_{AB}{}^ET_{EC}{}^D\Big)
~,
\label{Bianchi01-4}
\\
0&=&\sum_{[ABC)}\Big(\de_{A}(R_\cV)_{BC}{}^{kl}-T_{AB}{}^{D}(R_\cV)_{DC}{}^{kl}\Big)
~,
\label{Bianchi02-4}
\\
0&=&\sum_{[ABC)}\Big(\de_{A}(R_\cC)_{BC}{}^{kl}-T_{AB}{}^{D}(R_\cC)_{DC}{}^{kl}\Big)
~,
\label{Bianchi03-4}
\\
0&=&\sum_{[ABC)}\Big(\de_{A}R_{BC}-T_{AB}{}^{D}R_{DC}\Big)
~,
\label{Bianchi04-4}
\eea
\esubeq
where\footnote{In this appendix 
we often use the condensed notation 
$A_{\alu}\equiv A_{\a i}$ and $B_{\dot{\alu}}\equiv B_\a^i$; for instance we have 
$\de_{\alu}=\de_{\a i}$ and $\deb_{\dot{\alu}}=\deb_\a^i$.}
\bsubeq
\bea
&&R_{ABC}{}^D\equiv 
(R_\cV)_{AB}{}^{kl}(\cV_{kl})_C{}^D
+(R_\cC)_{AB}{}^{kl}(\cC_{kl})_C{}^D
+R_{AB}(\cM)_C{}^D
~,
\\
&&(\cV_{kl})_{{A}}{}^{{B}}\de_{{B}}\equiv {[}\cV_{kl},\de_{{A}}{]}~,~~
(\cC_{kl})_{{A}}{}^{{B}}\de_{{B}}\equiv {[}\cC_{kl},\de_{{A}}{]}~,~~
(\cm)_{{A}}{}^{{B}}\de_{{B}}\equiv {[}\cm,\de_{{A}}{]}~,
~~~~~~
\\
&&(\cV_{kl})_{\alu}{}^{\beu}=\hf\d_\a^\b C_{i(k}\d_{l)}^j~,~~~~~~
(\cV_{kl})_{\dot{\alu}}{}^{\dot{\beu}}=\hf\d_\a^\b\d^i_{(k}C_{l)j}~,
\\
&&(\cC_{kl})_{\alu}{}^{\beu}=\hf(\g^3)_\a{}^\b C_{i(k}\d_{l)}^j~,~~~~~~
(\cC_{kl})_{\dot{\alu}}{}^{\dot{\beu}}=\hf(\g^3)_\a{}^\b\d^i_{(k}C_{l)j}~,
\\
&&(\cm)_{\alu}{}^{\beu}=\hf\d_i^j (\g^3)_\a{}^\b~,~~~
(\cm)_{\dot{\alu}}{}^{\dot{\beu}}=\hf\d^i_j(\g^3)_{\a}{}^{\b}~,~~~
(\cm)_{{a}}{}^{{b}}=\ve_a{}^b~,
\eea
\esubeq
with the other components of $(\cV_{kl})_C{}^D$, $(\cC_{kl})_C{}^D$ and $(\cM)_C{}^D$
being equal to zero.

In solving the Bianchi identities it is important to remember that, 
due to Dragon's second theorem \cite{Dragon},
it is sufficient to analyze only eq. (\ref{Bianchi01-4});
all the equations (\ref{Bianchi02-4})--(\ref{Bianchi04-4}) are identically 
satisfied, provided that (\ref{Bianchi01-4}) holds.
This gives a great reduction in the number of equations that have to be studied.

To distinguish the curvatures and unambiguously raise and lower indices, we introduce the 
notation 
\bea
\bR_{\a i}{}_{\b j}:=R_{\alu\beu}~,~~~
\bar{\bR}_\a^i{}_\b^j:=R_{\dot{\alu}\dot{\beu}}~,~~~
\hat{\bR}_{\a i}{}_\b^j=R_\alu{}_{\dot{\beu}}~,~~~
\bR_{a\b j}:=R_{a\beu}~,~~~
\bar{\bR}_a{}_\b^j:=R_a{}_{\dot{\beu}}~,
\label{Rdef}
\eea
and similarly for the SU(2) curvatures.
Analogously, for the torsion it is useful to define different objects to freely raise and lower indices
\begin{subequations}
\bea
&{\cT}_{a}{}_{\b j}{}^{\g k}:=T_a{}_\beu{}^\gau~,~~~
\bar{\cT}_{a}{}_{\b}^{ j}{}^{\g}_{ k}:=T_a{}_{\dot{\beu}}{}^{\dot{\gau}}~,~~~
{\bT}_{a}{}_{\b j}{}^{\g}_k:=T_a{}_\beu{}^{\dot{\gau}}~,~~~
\bar{\bT}_{a}{}_{\b}^{j}{}^{\g k}:=T_a{}_{\dot{\beu}}{}^\gau~,
\label{Tdef1}
\\
&
\bT_{ab}{}^{\g k}:=T_{ab}{}^\gau~,~~~
\bar{\bT}_{ab}{}^\g_k:=T_{ab}{}^{\dot{\gau}}~.
\label{Tdef2}
\eea
\end{subequations}

At dimension-$1/2$, due to the choice  of the torsion constraints 
(\ref{constr-0})--(\ref{constr-1}),
the Bianchi identities are identically satisfied.
The non-trivial analysis begins at dimension-1.

\subsection{dimension-1}

${}$At dimension-1 there are many Bianchi identities that originate from eq 
 (\ref{Bianchi01-4}).
In fact, (\ref{Bianchi01-4}) gives the following set of equations: 
with (${A}=a,~{B}=\beu,~{C}=\gau,~{D}=d$)
\begin{subequations}
\bea
0&=&\bR_{\b j}{}_{\g k}\ve_{a}{}^d
+2\ri \bT_{a}{}_{\b j}{}^{\r}_q\,\d^{q}_{k}(\g^d)_{\g\r}
+2\ri \bT_a{}_{\g k}{}^{\r}_q\,\d^q_j(\g^d)_{\b\r}~,
\label{1-1-4}
\eea
with (${A}=a,~{B}={\beu},~{C}=\dot{\gau},~{D}=d$)
\bea
0&=&\hat{\bR}_{\b j}{}_{\g}^k\ve_{a}{}^d
+2\ri \cT_{a}{}_{\b j}{}^{\r q}\d_{q}^{k}(\g^d)_{\r\g}
+2\ri \bar{\cT}_a{}_{\g}^k{}^{\r}_q\d^q_j(\g^d)_{\b\r}~,
\label{1-2-4}
\eea
with (${A}=\alu,~{B}={\beu},~{C}=\dot{\gau},~{D}=\dot{\deu}$)
\bea
0&=&
(\bR_\cV)_{\a i}{}_{\b j}{}^k{}_l\d_\g^\d
+ (\bR_\cC)_{\a i}{}_{\b j}{}^k{}_l(\g^3)_\g{}^\d
+\hf \bR_{\a i}{}_{\b j}(\g^3)_\g{}^\d\d_l^k
\non\\
&&
+2\ri \d_j^k(\g^e)_{\b\g}\bT_{e}{}_{\a i}{}^{\d}_l
+2\ri \d^k_i(\g^e)_{\a\g}\bT_{e}{}_{\b j}{}^{\d}_l
~,
\label{1-3-4}
\eea
with (${A}=\alu,~{B}={\dot{\beu}},~{C}=\dot{\gau},~{D}=\dot{\deu}$)
\bea
0&=&
(\hat{\bR}_\cV)_{\a i}{}_{\b}^j{}^k{}_l\d^\d_\g
+(\hat{\bR}_\cC)_{\a i}{}_{\b}^j{}^k{}_l(\g^3)_\g{}^\d
+\hf \hat{\bR}_{\a i}{}_{\b}^j(\g^3)_\g{}^\d\d_l^k
\non\\
&&
+(\hat{\bR}_\cV)_{\a i}{}_{\g}^k{}^j{}_l\d_\b^\d
+(\hat{\bR}_\cC)_{\a i}{}_{\g}^k{}^j{}_l(\g^3)_\b{}^\d
+\hf \hat{\bR}_{\a i}{}_{\g}^k(\g^3)_\b{}^\d\d_l^j
\non\\
&&
+2\ri \d_i^j(\g^e)_{\a\b}\bar{\cT}_{e}{}_{\g}^k{}^{\d}_l
+2\ri \d^k_i(\g^e)_{\a\g}\bar{\cT}_{e}{}_{\b}^j{}^{\d}_l
~,\label{1-4-4}
\eea
and with (${A}={\alu},~{B}={{\beu}},~{C}={\gau},~{D}={\deu}$)
\bea
0&=&
(\bR_\cV)_{\a i}{}_{\b j}{}_k{}^l\d^\d_\g
+(\bR_\cC)_{\a i}{}_{\b j}{}_k{}^l(\g^3)_\g{}^\d
-\hf \bR_{\a i}{}_{\b j}(\g^3)_\g{}^\d\d^l_k
\non\\
&&
+(\bR_\cV)_{\b j}{}_{\g k}{}_i{}^l\d^\d_\a
+(\bR_\cC)_{\b j}{}_{\g k}{}_i{}^l(\g^3)_\a{}^\d
-\hf \bR_{\b j}{}_{\g k}(\g^3)_\a{}^\d\d^l_i
\non\\
&&
+(\bR_\cV)_{\g k}{}_{\a i}{}_j{}^l\d_\b^\d
+(\bR_\cC)_{\g k}{}_{\a i}{}_j{}^l(\g^3)_\b{}^\d
-\hf \bR_{\g k}{}_{\a i}(\g^3)_\b{}^\d\d^l_j
~.\label{1-5-4}
\eea
\end{subequations}
Here we have omitted identities that follow by complex conjugating the previous ones.

Equation (\ref{1-1-4}) gives the relation
\bea
\bR_{\b j}{}_{\g k}
&=&
\ri\ve^{ab} \bT_{a}{}_{\b j}{}^{\r}_k\,(\g_b)_{\g\r}
+\ri\ve^{ab} \bT_a{}_{\g k}{}^{\r}_j(\g_b)_{\b\r}~,
\label{1-1-1-4}
\eea
and also the following constraints to the torsion $\bT_{a}{}_{\b j}{}^\g_k$
\bea
0&=&
\bT_{(a}{}_{\b}^{j}{}_\r^{k}(\g_{b)})_{\g}{}^{\r}
+\bT_{(a}{}_{\g}^{k}{}_\r^{j}(\g_{b)})_{\b}{}^{\r}~.
\label{1-1-2-4}
\eea
These equations set to zero some irreducible components of the torsion and imply
\begin{subequations}
\bea
\bT_a{}_{\b j}{}_\g^k&=&
C_{\b\g}\d_j^{k}A_a
+\d_j^k(\g^3)_{\b\g}C_a
+\d_j^k\ve_{ab}(\g^b)_{\b\g}N
+(\g_a)_{\b\g}Y_j{}^{k}~,~~~
~~~~~~
\label{1-000-1}
\\
\bR_{\a i}{}_{\b j}
&=&
-4\ri C_{ij}C_{\a\b}N
+4\ri Y_{ij}(\g^3)_{\a\b}
~,
\label{1-000-2}
\eea
\end{subequations}
where the complex superfield $Y_{ij}$ is symmetric $Y_{ij}=Y_{ji}$.

Now, consider equation (\ref{1-3-4}) which, once used (\ref{1-000-1}) and
(\ref{1-000-2}), turns out to be equivalent to
\bea
&&(\bR_\cV)_{\a i}{}_{\b j}{}_{kl}C_{\g\d}
+(\bR_\cC)_{\a i}{}_{\b j}{}_{kl}(\g^3)_{\g\d}=
-2\ri C_{ij}C_{kl}C_{\a\b}(\g^3)_{\g\d} N
-2\ri C_{jk}C_{il}(\g^a)_{\b\g}\ve_{ab}(\g^b)_{\a\d}N
\non\\
&&~~~
-2\ri C_{ik}C_{jl}(\g^a)_{\a\g}\ve_{ab}(\g^b)_{\b\d}N
+2\ri (\g^3)_{\a\b}(\g^3)_{\g\d} C_{kl}Y_{ij}
-2\ri C_{jk}(\g^a)_{\b\g}(\g_a)_{\a\d}Y_{il}
\non\\
&&~~~
-2\ri C_{ik}(\g^a)_{\a\g}(\g_a)_{\b\d}Y_{jl}
-2\ri C_{jk}(\g^a)_{\b\g}C_{\a\d}C_{il}A_a
-2\ri C_{ik}(\g^a)_{\a\g}C_{\b\d}C_{jl}A_a
\non\\
&&~~~
-2\ri C_{jk}(\g^a)_{\b\g}C_{il}(\g^3)_{\a\d}C_a
-2\ri C_{ik}(\g^a)_{\a\g}C_{jl}(\g^3)_{\b\d}C_a
~.
\label{eq-1-1-4}
\eea
Taking the trace of the previous equation with $(\g^3)^{\g\d}$ one finds
\bea
(\bR_\cC)_{\a i}{}_{\b j}{}^{kl}&=&
\ri (\g^3)_{\a\b}\big(\d_{i}^{(k}Y_{j}{}^{l)}
+\d_{j}^{(k}Y_{i}{}^{l)}\big)
-\ri  (\g_a)_{\a\b}\,\d_{i}^{(k}\d_{j}^{l)}\big(\ve^{ab}A_b+C^a\big)
~.~~~~~~~~~
\eea
Taking the trace of equation (\ref{eq-1-1-4}) with $C^{\g\d}$ the following equation follows
\bea
(\bR_\cV)_{\a i}{}_{\b j}{}^{kl}&=&
2\ri \d_{i}^{(k}\d_{j}^{l)}(\g^3)_{\a\b}N
+2\ri C_{ij}C_{\a\b}Y^{kl}
-\ri \d_{i}^{(k}\d_{j}^{l)}(\g^a)_{\a\b}\big(A_a+\ve_{ab}C^b\big)
~.~~~~~~~~~
\eea
Then, from the trace of equation (\ref{eq-1-1-4}) with $(\g^c)^{\g\d}$ one finds the constraint
\bea
C_a=\ve_{ab}A^{b}~.
\eea
Summarizing the results obtained so far, it holds
\begin{subequations}
\bea
\bT_a{}_{\b j}{}_\g^k&=&
\d_j^{k}C_{\b\g}A_a
+\d_j^k(\g^3)_{\b\g}\ve_{ab}A^b
+\d_j^k\ve_{ab}(\g^b)_{\b\g}N
+(\g_a)_{\b\g}Y_j{}^{k}
~,
\label{T-1111}
\\
\bR_{\a i}{}_{\b j}
&=&
-4\ri C_{ij}C_{\a\b}N
+4\ri Y_{ij}(\g^3)_{\a\b}
~,
\label{R-111}
\\
(\bR_\cV)_{\a i}{}_{\b j}{}^{kl}&=&
2\ri \d_{i}^{(k}\d_{j}^{l)}(\g^3)_{\a\b}N
+2\ri C_{ij}C_{\a\b}Y^{kl}
-2\ri \d_{i}^{(k}\d_{j}^{l)}(\g^a)_{\a\b}A_a
~,~~~~~~~~~
\label{RV-111}
\\
(\bR_\cC)_{\a i}{}_{\b j}{}^{kl}&=&
\ri (\g^3)_{\a\b}\big(\d_{i}^{(k}Y_{j}{}^{l)}
+\d_{j}^{(k}Y_{i}{}^{l)}\big)
-2\ri  \d_{i}^{(k}\d_{j}^{l)}(\g_a)_{\a\b}\ve^{ab}A_b
~.~~~~~~~~~
\label{RC-111}
\eea
\end{subequations}

Let us now consider the Bianchi identity (\ref{1-2-4})
which is equivalent to
\bea
\hat{\bR}_{\b j}{}_{\g k}\ve_{ab}&=&
2\ri \cT_{a}{}_{\b j}{}_{\r k}(\g_b)_{\g}{}^\r
+2\ri \bar{\cT}_a{}_{\g k}{}_{\r j}(\g_b)_{\b}{}^{\r}~.
\label{V-1-1-1-4}
\eea
Note that, by considering a real superfield  
$(A)^*=A~,\ve(A)=0~,(\de_{\a i} A)^*=-\deb_{\a}^iA$, being
\bea
({[}\de_a,\de_{\a i}{]}A)^*
=-(\cT_a{}_{\a i}{}^{\b j})^*\deb_{\b}^jA+\cdots~,~~~
({[}\de_a,\de_{\a i}{]}A)^*
=-\bar{\cT}_a{}_\a^i{}^\b_j\deb_{\b}^jA+\cdots
~,
\eea
we obtain the complex conjugation relation between $\cT_a{}_\b^j{}^\g_k$ and 
$\bar{\cT}_a{}_{\b j}{}^{\g k}$:
\bea
\bar{\cT}_a{}_\b^j{}^\g_k=(\cT_a{}_{\b j}{}^{\g k})^*~.
\label{compl-111}
\eea
Using the torsion constraint (\ref{constr-1}) together with (\ref{compl-111}),
the symmetric part in $a,b$ of eq. (\ref{V-1-1-1-4}) implies that the torsion $\cT_a{}_{\a i}{}^{\b j}$
is
\bea
\cT_a{}_{\b j}{}^{\g k}
&=&
(\g_a)_{\b}{}^{\g}\big(\ri\d_j^k\cS+\cS_j{}^k\big)
+\ve_{ab}(\g^b)_{\b}{}^{\g}\big(\d_j^k\cT+\ri\cT_j{}^k\big)
\non\\
&&
+\ri\d_\b^\g\d_j^k\cB_a
+\ri(\g^3)_{\b}{}^{\g}\d_j^k\cC_a
~,~~~
\label{T-111111}
\eea
where
\begin{subequations}
\bea
&(\cS)^*=\cS~,~~(\cT)^*=\cT~,~~(\cB_a)^*=\cB_a~,~~(\cC_a)^*=\cC_a~,
\\
&(\cS^{ij})^*=\cS_{ij}~,~~(\cT^{ij})^*=\cT_{ij}~,~~~
\cS_{ij}=\cS_{ji}~,~~\cT_{ij}=\cT_{ji}~.
\eea
\end{subequations}
The antisymmetric part in $a,b$ of (\ref{V-1-1-1-4}) is solved by 
\bea
\hat{\bR}_{\a i}{}_{\b}^{j}&=&
-4\ri C_{\a\b}\big(\d_{i}^{j}\cT+\ri\cT_{i}{}^{j}\big)
+4\ri(\g^3)_{\a\b}\big(\ri \d_{i}^{j}\cS+\cS_{i}{}^{j}\big)
~.
\label{R-111111}
\eea

Let us now turn our attention to eq. (\ref{1-4-4}) which is equivalently written as 
\bea
&&(\hat{\bR}_\cV)_{\a}^i{}_{\b}^j{}^{kl}C_{\g\d}
+(\hat{\bR}_\cC)_{\a}^i{}_{\b}^j{}^{kl}(\g^3)_{\g\d}
+(\hat{\bR}_\cV)_{\a}^i{}_{\g}^k{}^{jl}C_{\b\d}
+(\hat{\bR}_\cC)_{\a}^i{}_{\g}^k{}^{jl}(\g^3)_{\b\d}=
~~~~~~
\non\\
&&
~~~=\hf \hat{\bR}_{\a}^i{}_{\b}^j(\g^3)_{\g\d} C^{kl}
+\hf \hat{\bR}_{\a}^i{}_{\g}^k(\g^3)_{\b\d} C^{jl}
-2\ri C^{ij}(\g^e)_{\a\b}\bar{\cT}_{e}{}_{\g}^k{}_{\d}^l
-2\ri C^{ik}(\g^e)_{\a\g}\bar{\cT}_{e}{}_{\b}^j{}_{\d}^l
~.~~~~~~
\label{VC-1-1}
\eea
The right hand side of the previous equation can be expressed in terms of the torsion 
components $\cS,\cT,\cB_a,\cC_a,\cS_{ij}$ and $\cT_{ij}$ by making use of the equations
(\ref{T-111111}), (\ref{compl-111}) and (\ref{R-111111}). The solution of eq. (\ref{VC-1-1})
can then be approached by considering the trace of (\ref{VC-1-1}) with 
$C^{\g\d}$, $(\g^3)^{\g\d}$ and $(\g_c)^{\g\d}$ respectively. 
Solving the three resulting equations
 one finds a new constraint for the torsion components
\bea
\cC_a=\ve_{ab}\cB^b~,
\label{BA-111}
\eea
and also the following expressions for the remaining SU(2) curvatures
\begin{subequations}
\bea
(\hat{\bR}_\cV)_{\a i}{}_{\b}^j{}^{kl}
&=&
-2\ri(\g^3)_{\a\b}\d_i^{(k}C^{l)j}\cT
-2C_{\a\b}\d_i^{(k}C^{l)j}\cS
+2(\g^3)_{\a\b}\d_i^j\cT^{kl}
+2\ri C_{\a\b}\d_i^j\cS^{kl}
\non\\
&&
-2\d_i^{(k}C^{l)j}(\g_a)_{\a\b}\cB^a
\label{RV-111111}
~,
\\
(\hat{\bR}_\cC)_{\a i}{}_{\b}^{j}{}^{kl}&=&
C_{\a\b}\Big(
\d_i^{(k}\cT^{l)j}
+C^{j(k}\cT^{l)}{}_i
\Big)
+(\g^3)_{\a\b}\Big(
\ri \d_i^{(k}\cS^{l)j}
+\ri C^{j(k}\cS^{l)}{}_i
\Big)
\non\\
&&
-2\d_i^{(k}C^{l)j}\ve_{ab}(\g^a)_{\a\b}\cB^b
\label{RC-111111}
~.
\eea
\end{subequations}
Note that (\ref{BA-111}) simplifies the torsions $\cT_a{}_{\b j}{}^{\g k}$ and 
$\bar{\cT}_a{}_\b^j{}^\g_k$
to their final form
\begin{subequations}\bea
\cT_a{}_{\b j}{}^{\g k}&=&
\ve_{ab}(\g^b)_{\b}{}^{\g}\big(\d_{j}^k\cT+\ri\cT_{j}{}^{k}\big)
+(\g_a)_{\b}{}^{\g}\big(\ri \d_j^k\cS+\cS_{j}{}^{k}\big)
\non\\
&&
+\ri\d_\b^\g\d_j^k\cB_a
+\ri(\g^3)_{\b\g}\d_j^k\ve_{ab}\cB^b
~,~~~
\label{T-11111}
\\
\bar{\cT}_a{}_{\b}^j{}^{\g}_k&=&
-\ve_{ab}(\g^b)_{\b}{}^{\g}\big(\d_k^j\cT+\ri\cT^{j}{}_k\big)
+(\g_a)_{\b}{}^{\g}\big(\ri \d_k^j\cS+\cS^{j}{}_k\big)
\non\\
&&
-\ri\d_{\b}^\g\d_k^j\cB_a
-\ri(\g^3)_{\b}{}^{\g}\d_k^j\ve_{ab}\cB^b
~.
\label{T-1111112}
\eea
\end{subequations}

By making use of the expressions obtained for the curvatures (\ref{R-111})--(\ref{RC-111}), 
the last Bianchi identity to be checked, eq. (\ref{1-5-4}), turns out to be identically satisfied.
This concludes the analysis of the  dimension-1 Bianchi
 identities since other Bianchi identities, not explicitly studied,
 are identically satisfied by taking into account complex conjugation.

\subsection{dimension-3/2}

We begin the analysis of the dimension-3/2 Bianchi identities by considering
eq. (\ref{Bianchi01-4}) with (${A}=a,~{B}=\beu,~{C}=\gau,~{D}=\dot{\deu}$)
\bea
0&=&
-\de_{\b j}\bT_{a\g k}{}^{\d}_l
-\de_{\g k}\bT_{a\b j}{}^{\d}_l
~.
\eea
From the previous equation, a set of dimension-3/2 
differential constraints on the torsion components $N,\,Y_{ij}$ and $A_a$ arises:
\begin{subequations}
\bea
\de_{\a}^iY^{jk}&=&(\g^3)_{\a}{}^{\b}C^{i(j}\de_{\b}^{k)}N~,
\\
\de_{\b}^jA_a
&=&
-\ve_{ab}(\g^b)_{\b}{}^{\d}\de_{\d}^jN
~.
\eea
\end{subequations}

To continue, let us consider the Bianchi identity given by (\ref{Bianchi01-4}) 
with indices
(${A}=a,~{B}=b,~{C}={\gau},~{D}=d$):
\bea
0&=&
-\bR_{b}{}_{\g k}\ve_a{}^d
+\bR_a{}_{\g k}\ve_b{}^d
-2\ri \bar{\bT}_{ab}{}^{\rho}_p\d^p_k(\g^d)_{\g\rho}
~.
\eea
This is equivalent to the following equation for the dimension-3/2 Lorentz curvature
\bea
\bR_{a}{}_{\b j}
=\ri\ve^{bc}\bar{\bT}_{bc}{}^{\g}_j(\g_a)_{\g\b}~.
\label{R-3/2}
\eea

The Bianchi identity (\ref{Bianchi01-4}) with indices
(${A}=a,~{B}=\beu,~{C}=\dot{\gau},~{D}=\dot{\deu}$) is:
\bea
0&=&
-\d_\g^\d (\bR_\cV)_{a}{}_{\b j}{}^k{}_l
-(\g^3)_\g{}^\d (\bR_\cC)_{a}{}_{\b j}{}^k{}_l
-\hf\d^k_l\bR_a{}_{\b j}(\g^3)_\g{}^\d
-2\ri\d_j^k(\g^e)_{\b\g} \bar{\bT}_{ea}{}^\d_l
\non\\
&&
-\de_{\b j}\bar{\cT}_a{}_\g^k{}^\d_l
-\deb_\g^k\bT_{a}{}_{\b j}{}^\d_l
~.
\label{3/2-3-1}
\eea
By taking the trace of (\ref{3/2-3-1}) 
with $(\g^e)_\d{}^{\g}$ and solving the resulting 
equation one finds a set of constraints on the torsion components
\begin{subequations}
\bea
\deb_\b^{(j}Y^{kl)}
&=&
-2\de_{\b}^{(j}\cS^{kl)}
=
-2\ri(\g^3)_\b{}^{\g}\de_{\g}^{(j}\cT^{kl)}
~,
\label{deS-3/2-111}
\\
\de_{\b}^{j}\cS
&=&
{\ri\over 2}(\g^3)_\b{}^{\d}\deb_\d^{j}N
+{1\over 3}(\g^3)_\b{}^\g\de_{\g k}\cT^{jk}
-{\ri\over 6}\deb_{\b k}Y^{jk}
~,
\label{3/2-3-1-4-2}
\\
\de_{\b}^j\cT
&=&
-\hf\deb_\b^{j}N
+{1\over 3}(\g^3)_\b{}^{\d}\de_{\d k}\cS^{jk}
+{1\over 6}(\g^3)_\b{}^{\g}\deb_{\g k} Y^{jk}
~,\\
(\g^a)_\b{}^{\g}\deb_\g^{j}A_a
&=&
-{2\over 3}\de_{\b k}\cS^{jk}
+{2\ri\over 3}(\g^3)_\b{}^{\d}\de_{\d k}\cT^{jk}
~,
\label{3/2-111111}
\eea
\end{subequations}
and the following expression for the dimension-3/2 torsion 
\bea
\bar{\bT}_{ab}{}_\g^k
&=&
-\hf\ve_{ab}\Big(
\ri\deb_\g^{k}N
-{2\ri\over 3}(\g^3)_\g{}^{\d}\de_{\d l}\cS^{kl}
+{2\over 3}\de_{\g l}\cT^{kl}
+{\ri\over 3}(\g^3)_\g{}^{\d}\deb_{\d l}Y^{kl}
\Big)
~.
\label{Tt-3/2}
\eea
It is also useful to decompose $\deb_\g^kA_a$ in its irreducible gamma and gamma-traceless 
parts
\bea
\deb_\b^jA_a=A_{a}{}_\b^j+\hf(\g_a)_\b{}^\g(\g^b)_\g{}^\d\deb_\d^jA_b~,~~~~~~
(\g^a)_\a{}^\g A_{a}{}_\g^k=0~,
\eea
which by using (\ref{3/2-111111}) gives
\bea
\deb_\g^kA_a=A_{a}{}_\g^k
-{1\over 3}(\g_a)_\g{}^\d\de_{\d l}\cS^{kl}
+{\ri\over 3}\ve_{ab}(\g^b)_\g{}^\d\de_{\d l}\cT^{kl}
~.
\label{cA-3/2}
\eea
By taking the trace of eq. (\ref{3/2-3-1}) with $\d_\d^{\g}$  and by using eqs. 
(\ref{deS-3/2-111})--(\ref{cA-3/2})
we obtain the constraint
\bea
 \de_{\b}^{j}\cB_a
&=&
{\ri\over 6}(\g_a)_{\b}{}^{\g}\de_{\g p}\cS^{jp}
-{1\over 6}\ve_{ab}(\g^b)_{\b}{}^{\g}\de_{\g p}\cT^{jp}
+{\ri\over 6}(\g_a)_{\b}{}^{\g}\deb_{\g p}Y^{jp}
+{\ri\over 2}A_a{}_\b^j
~,~~~~~~~~~
\eea
and the following expression for the dimension-3/2 SU(2)$_\cV$ curvature
\bea
 (\bR_\cV)_{a}{}_{\b j}{}^{kl}
&=&
-{1\over 2} \d_j^{(k}\ve_{ab}(\g^b)_{\b}{}^{\g}\deb_\g^{l)}N
+{1\over 6}\d_j^{(k}(\g_a)_{\b}{}^{\g}\de_{\g p}\cS^{l)p}
+{\ri\over 6}\d_j^{(k}\ve_{ab}(\g^b)_{\b}{}^{\g}\de_{\g p}\cT^{l)p}
\non\\
&&
-{1\over 12} (\g_a)_{\b}{}^{\g}\deb_\g^{(p}Y^{kl)}C_{pj}
-{1\over 2} \d_j^{(k}A_a{}_\b^{l)}
~.~~~~~~~~~~~~
\label{RV-3/2}
\eea
To complete the analysis of the Bianchi identity (\ref{3/2-3-1}) it is necessary to analyze its
trace with $(\g^3)_\d{}^{\g}$. Solving the resulting equation, with the help of the results obtained 
so far, the following expression for the dimension-3/2 SU(2)$_\cC$ curvature arises
\bea
(\bR_\cC)_{a}{}_{\b j}{}^{kl}
&=&
-{1\over 6}\d_j^{(k}\ve_{ab}(\g^b)_{\b}{}^{\g}\de_{\g p}\cS^{l)p}
-{\ri\over 6}\d_j^{(k}(\g_a)_{\b}{}^{\d}\de_{\d p}\cT^{l)p}
+{1\over 6} \d_j^{(k}\ve_{ab}(\g^b)_{\b}{}^{\d}\deb_{\d p}Y^{l)p}
\non\\
&&
-{1\over 12} \ve_{ab}(\g^b)_{\b}{}^{\d}\deb_\d^{(p}Y^{kl)}C_{pj}
-{1\over 2} \d_j^{(k}\ve_{ab}A^b{}_\b^{l)}
~.
\label{RC-3/2}
\eea

It remains to consider the Bianchi identity (\ref{Bianchi01-4}) with
(${A}=a,~{B}=\beu,~{C}=\gau,~{D}={\deu}$):
\bea
0&=&
(\bR_\cV)_a{}_{\b j}{}_{k}{}^{l}\d_\g^\d
+(\bR_\cV)_a{}_{\g k}{}_{j}{}^l\d_\b^\d
+(\bR_\cC)_a{}_{\b j}{}_{k}{}^l(\g^3)_\g{}^\d
+(\bR_\cC)_a{}_{\g k}{}_{j}{}^l(\g^3)_\b{}^\d
\non\\
&&
-\hf \bR_a{}_{\b j}(\g^3)_\g{}^\d\d_k^l
-\hf \bR_a{}_{\g k}(\g^3)_\b{}^\d\d_j^l
-\de_{\b j}\cT_a{}_{\g k}{}^{\d l} 
-\de_{\g k}\cT_a{}_{\b j}{}^{\d l} 
~.
\eea
This turns out to be identically satisfied once used the results previously obtained.
The rest of the dimension-3/2 Bianchi identities, not explicitly written here, are satisfied by taking 
into account complex conjugation.


\subsection{dimension-2}
\label{dim-2}

At dimension-2 the Bianchi identity (\ref{Bianchi01-4}) with $(A=a,B=b,C=\gau,D=\deu)$ 
gives
\bea
0&=&\ve^{ab}(R_\cV)_{ab}{}_k{}^{l}\d_\g^\d
+\ve^{ab}(R_\cC)_{ab}{}_k{}^{l}(\g^3)_{\g}{}^{\d}
-\hf \ve^{ab}R_{ab}(\g^3)_{\g}{}^{\d}\d_k^l
+\de_{\g k}\ve^{ab}\bT_{ab}{}^{\d l}
+2\ve^{ab}\de_{a}\cT_{b\g k}{}^{\d l}
\non\\
&&
-2\ve^{ab}\cT_{a\g k}{}^{\a i}\cT_{b\a i}{}^{\d l}
-2\ve^{ab}\bT_{a\g k}{}^{\a}_i\bar{\bT}_{b}{}_\a^i{}^{\d l}
~.
\label{2-0}
\eea
One can first consider the trace of the previous equation with $(\g_c)_\d{}^{\g}$. Such equation 
turns out to be identically satisfied.
To prove it one may use the following identities 
 \begin{subequations}
 \bea
\ri(\g_a)^{\a\b}{[}\de_{\a i},\de_{\b}^{i}{]}\bar{N}
&=&
-{2\ri\over 3}\ve_{ab}(\g^b)^{\a\b}{[}\de_{\a i},\deb_{\b j}{]}\cS^{ij}
+{2\over 3}(\g_a)^{\a\b}{[}\de_{\a i},\deb_{\b j}{]}\cT^{ij}
\non\\
&&
+16\de_{a}\cT
+16\ri\ve_{ab}\de^{b}\cS
-16 A_a\bar{N}
~,~~~~~~~~~
\label{DDT-2-111}
\\
\ri(\g_a)^{\a\b}{[}\deb_{\a i},\deb_\b^{i}{]}N
&=&
-{2\ri\over 3}\ve_{ab}(\g^b)^{\a\b}{[}\de_{\a i},\deb_{\b j}{]}\cS^{ij}
-{2\over 3}(\g_a)^{\a\b}{[}\de_{\a i},\deb_{\b j}{]}\cT^{ij}
\non\\
&&
+16\de_{a}\cT
-16\ri\ve_{ab}\de^{b}\cS
-16 \bar{A}_aN
~,~~~~~~~~~
\\
(\g^a)^{\a\b}\de_\d^{(i}\deb_{\g k}\cT^{jk)}
&=&
32\ri\de^a\cT^{ij}
+{2\ri} \ve^{ab}(\g_b)^{\a\b}\deb_{\b}^{(i}\deb_{\a p}Y^{j)p}
+64\cB^a\cT^{ij}
-48\ve^{ab}\bar{A}_bY^{ij}
~,~~~~~~~~~
\\
(\g^a)^{\a\b}\de_\a^{(i}\deb_{\b k}\cS^{jk)}
&=&
32\ri\de^a\cS^{ij}
-2(\g^a)^{\a\b}\deb_{\a}^{(i}\deb_{\b p}Y^{j)p}
+64\cB^a\cS^{ij}
-{48\ri} \bar{A}^a{Y}^{ij}
~,
\label{DDS-2-111}
\eea
 \end{subequations}
where clearly $\bar{N}=(N)^*$ and $\bar{Y}_{ij}=(Y^{ij})^*$.
 These equations are consequences of the dimension-3/2 Bianchi identities
 (\ref{3/2-a})--(\ref{3/2-g}). One can obtain (\ref{DDT-2-111})--(\ref{DDS-2-111})  in two steps:  
 first derive a set of dimension-2 
 differential equations by applying a spinor covariant derivative to eqs. (\ref{3/2-a})--(\ref{3/2-g}); 
 then manipulate the resulting equations by taking into account the dimension-1 covariant 
 derivatives  algebra (\ref{Algebra-1.1})--(\ref{Algebra-1.1c}) and the structure group 
 transformation properties of the dimension-1 torsion components (\ref{structT1})--(\ref{structT3}).

Now, consider eq. (\ref{2-0}) contracted with 
$\d_\d^{\g}\d_{l}^k$. The resulting equation is identically satisfied by making use of 
\bea
(\g^3)^{\a\g}{[}\de_{\a i},\deb_{\g j}{]}\cS^{ij}
-\ri{[}\de_{\a i},\deb^\a_{j}{]}\cT^{ij}
+24\ve^{ab}\de_a\cB_b
=0
\eea
which is again a dimension-2 consequence of (\ref{3/2-a})--(\ref{3/2-g}).

Contracting equation (\ref{2-0}) with $(\g^3)_{\d}{}^\g \d^k_{l}$ the dimension-2 Lorentz curvature 
$R_{ab}$ can be computed.
With the aid of the equations
\bsubeq
\bea
0&=&{[}\de_{\a i},\de^\a_{j}{]}\bar{Y}^{ij}
+{[}\deb_{\a i},\deb^\a_{ j}{]}Y^{ij}
-24\de^a\cB_a
~,
\label{2-111111-1}
\\
0&=&
\ri(\g^3)^{\a\b}{[}\deb_{\a i},\deb_{\b}^i{]}N
+\ri(\g^3)^{\a\b}{[}\de_{\a i},\de_{\b}^i{]}\bar{N}
-{\ri\over 3}{[}\deb_{\a i},\deb^\a_{j}{]}Y^{ij}
-{\ri\over 3}{[}\de_{\a i},\de^\a_{j}{]}\bar{Y}^{ij}
~,~~~~~~~~~
\label{2-111111-2}
\eea
\esubeq
we find the expression
\bea
 R_{ab}
&=&
-\hf\ve_{ab}\Big(\,
{\ri\over 4}(\g^3)^{\a\b}{[}\deb_{\a i},\deb_{\b}^i{]}N
-{\ri\over 4}(\g^3)^{\a\b}{[}\de_{\a i},\de_{\b}^i{]}\bar{N}
+{\ri\over 12}{[}\deb_{\a i},\deb^\a_{j}{]}Y^{ij}
-{\ri\over 12}{[}\de_{\a i},\de^\a_{j}{]}\bar{Y}^{ij}
\non\\
&&~~~~~~~~~
-{\ri\over 6}{[}\de_{\a i},\deb^\a_{j}{]}\cS^{ij}
-{1\over 6}(\g^3)^{\a\b}{[}\de_{\a i},\deb_{\b j}{]}\cT^{ij}
+8\cT^2
+8\bar{N}N
+8\cS^2
\non\\
&&~~~~~~~~~
+4\cS^{ij}\cS_{ij}
+4\cT^{ij}\cT_{ij}
+4\bar{Y}^{ij}Y_{ij}
\,\Big)
~.~~~~~~~~~
\label{2-R-Lorentz-3}
\eea
Note that the relations (\ref{2-111111-1}) and (\ref{2-111111-2}) 
again derive from (\ref{3/2-a})--(\ref{3/2-g}).

As a next step, contract equation (\ref{2-0}) with $\d_{\d}^\g$ and take the traceless part in the 
SU(2) indices $k,l$. From the resulting equation, by also using the relations
\bsubeq
\bea
0&=&
{[}\de_{\a (i},\de^\a_{j)}{]}\bar{N}
+{[}\deb_{\a (i},\deb^\a_{j)}{]}N
+(\g^3)^{\a\b}{[}\de_{\a p},\de_{\b}^p{]}\bar{Y}_{ij}
\non\\
&&
+(\g^3)^{\a\b}{[}\deb_{\a p},\deb_{\b}^p{]}Y_{ij}
~,
\\
\deb_{\a p}\de^{\a p}\cT_{ij}
&=&
{\ri\over 8}(\g^3)^{\a\b}{[}\deb_{\a p},\deb_{\b}^p{]}{Y}_{ij}
-{\ri\over 8}(\g^3)^{\a\b}{[}\de_{\a p},\de_{\b}^p{]}\bar{Y}_{ij}
-4\ri\cS_{(i}{}^k\cT_{j)k}
~,~~~~~~
\\
(\g^3)^{\a\b}\deb_{\a p}\de_{\b}^p\cS_{ij}
&=&
{1\over 8}(\g^3)^{\a\b}{[}\de_{\a p},\de_{\b}^p{]}\bar{Y}_{ij}
-{1\over 8}(\g^3)^{\a\b}{[}\deb_{\a p},\deb_{\b}^p{]}Y_{ij}
+4\cS_{(i}{}^k\cT_{j)k}
~,
\eea
\esubeq
that derive from (\ref{3/2-a})--(\ref{3/2-g}),
we find 
the dimension-2 SU(2)$_\cV$ curvature
\bea
(R_\cV)_{ab}{}^{kl}
&=&
-\hf\ve_{ab}\Big(\,
{\ri\over 16}{[}\deb_{\a}^{(k},\deb^{\a l)}{]}N
-{\ri\over 16}{[}\de_{\a}^{(k},\de^{\a l)}{]}\bar{N}
-{\ri\over 16}(\g^3)^{\a\b}{[}\deb_{\a p},\deb_{\b}^{p}{]}Y^{kl}
\non\\
&&~~~~~~~~~
+{\ri\over 16}(\g^3)^{\a\b}{[}\de_{\a p},\de_{\b}^p{]}\bar{Y}^{kl}
+8\cS^{kl}\cT
+8\ri\cS^{(k}{}_p\cT^{l)p}
\,\Big)
~.
\label{R-SU2-V-2}
\eea

To conclude the solution of eq. (\ref{2-0}),
we take its trace with $(\g^3)_\d{}^{\g}$ and consider the traceless part in the
SU(2) indices $k,l$.
By using (\ref{3/2-a})--(\ref{3/2-g}), that imply the relations\bsubeq
\bea
\deb_{\a k}\de^{\a k}\cS_{ij}
&=&
{1\over 16}{[}\deb_{\a (i},\deb^{\a k}{]}Y_{j)k}
+{1\over 16}{[}\de_{\a (i},\de^{\a k}{]}\bar{Y}_{j)k}
\non\\
&&
-2\ri \bar{Y}_{(i}{}^{k}Y_{j)k}
-8\cS\cS_{ij}
+8\cT\cT_{ij}
~,
\\
(\g^3)^{\a\b}\deb_{\b p}\de_{\a}^p\cT_{ij}
&=&
-{\ri\over 16}{[}\deb_{\a (i},\deb^{\a k}{]}Y_{j)k}
-{\ri\over 16}{[}\de_{\a (i},\de^{\a k}{]}\bar{Y}_{j)k}
\non\\
&&
-2\bar{Y}_{(i}{}^kY_{i)k}
+8\ri\cS\cS_{ij}
-8\ri\cT\cT_{ij}
~,
\eea
\esubeq
we  obtain the following expression for the dimension-2 SU(2)$_\cC$ curvature
\bea
(R_\cC)_{ab}{}^{kl}
&=&
-\hf\ve_{ab}\Big(
{\ri\over 48}{[}\deb^{\a(k},\deb_{\a p}{]}Y^{l)p}
-{\ri\over 48}{[}\de^{\a(k},\de_{\a p}{]}\bar{Y}^{l)p}
-4\bar{Y}^{p(k}Y_p{}^{l)}
\Big)
~.~~~~~~
\label{R-SU2-C-2}
\eea

The  Bianchi identity (\ref{Bianchi01-4}) with 
$(A=a,B=b,C=\gau,D=\dot{\deu})$ is equivalent to the equation
\bea
0&=&
\de_{\g k}\ve^{ab}\bar{\bT}_{ab}{}^{\d}_{l}
+2\ve^{ab}\de_{a}\bT_{b\g k}{}^{\d}_{l}
-2\ve^{ab}\cT_{a\g k}{}^{\a i}\bT_{b\a i}{}^{\d}_{l}
-2\ve^{ab}\bT_{a\g k}{}^{\a}_i\bar{\cT}_{b}{}_\a^i{}^{\d}_{l}
~.
\eea
This is identically satisfied by using  
(\ref{3/2-a})--(\ref{3/2-g}), (\ref{Algebra-1.1})--(\ref{Algebra-1.1c}) and 
(\ref{structT1})--(\ref{structT3}).
The rest of the dimension-2 Bianchi identities, not explicitly written here, are satisfied by 
taking into account complex conjugation.


\section{Derivation of eq. (\ref{dedeU})}
\setcounter{equation}{0}

In this appendix we give a derivation of eq. (\ref{dedeU})
which is crucial for the analysis of section \ref{projSuper}.
Consider a general weight-(m,n)  bi-isotwistor superfield $U^{(m,n)}$ as defined 
in subsection \ref{cov2-proj}.
First, we analyze the pure left sector of (\ref{dedeU}).
Being
\bea
(\g^0)_{++}=(\g^1)_{++}=(\g^0)_{--}=1~,~~
(\g^1)_{--}=-1~,~~~~
(\g^0)_{+-}=(\g^1)_{+-}=0~,
\eea
by using (\ref{Algebra-1.1}), one 
finds the $\{\de_+^\opl,\de_+^\opl\}$ spinor derivatives anticommutator
\bea
\{\de_{+}^\opl,\de_{+}^\opl\}&=&
-8\ri A_{++}\bmL^{\opl\opl}
~,
\eea
where $A_{++}=(\g^a)_{++}A_a=(A_0+A_1)$ and $\bmL^{\opl\opl}=u_i^\opl u_j^\opl\bmL^{ij}$.
From equation (\ref{LU}) it is easy to observe that it holds
\bea
\bmL^{\opl\opl}U^{(m,n)}=0~,~~~\Longrightarrow~~~
\{\de_{+}^\opl,\de_{+}^\opl\}U^{(m,n)}=0~.
\eea
In complete similarity, from (\ref{Algebra-1.2}), (\ref{Algebra-1.1c}) and 
$\bmL^{\opl\opl}U^{(m,n)}=0$ 
one finds
\bsubeq\bea
&\{\de_{+}^\opl,\deb_+^\opl\}=
8\cB_{++}\bmL^{\opl\opl}
~,
~~~
\{\deb_+^\opl,\deb_+^\opl\}=
-8\ri\bar{A}_{++}\bmL^{\opl\opl}
~,
\\
&\{\de_{+}^\opl,\deb_{+}^\opl\}U^{(m,n)}
=\{\deb_{+}^\opl,\deb_{+}^\opl\}U^{(m,n)}=0~.
\eea
\esubeq

The same works in the right light-cone sector of the algebra. In fact, with 
$A_{--}=(\g^a)_{--}A_a=(A_0-A_1)$ and $\bmR^{\bpl\bpl}=v_I^{\bpl}v_J^\bpl\bmR^{IJ}$
it holds
\bea
\{\de_{-}^\bpl,\de_{-}^\bpl\}=
-8\ri A_{--}\bmR^{\bpl\bpl}
~,~~
\{\de_{-}^\bpl,\deb_-^\bpl\}=
8\cB_{--}\bmR^{\bpl\bpl}
~,~~
\{\deb_-^\bpl,\deb_-^\bpl\}=
-8\ri\bar{A}_{--}\bmR^{\bpl\bpl}
\,,~~~~
\eea
and, being $\bmR^{\bpl\bpl}U^{(m,n)}=0$, one finds
\bea
\{\de_{-}^\bpl,\de_{-}^\bpl\}U^{(m,n)}
=\{\de_-^\bpl,\deb_{-}^\bpl\}U^{(m,n)}
=\{\deb_{-}^\bpl,\deb_{-}^\bpl\}U^{(m,n)}
=0
~.
\eea

Consider now the mixed left-right sector which results a bit less trivial.
Using
\bea
C_{+-}=-C_{-+}=-\ri~,~~~
(\g^3)_{+-}=(\g^3)_{-+}=-\ri~,
\eea
and (\ref{Algebra-1.1}), one finds 
\bea
\{\de_{+}^\opl,\de_{-}^\bpl\}&=&
4\big((u^\opl v^\bpl)N
+Y^{\opl\bpl}\big)\cm
+4N\cV^{\opl\bpl}
-2(u^\opl v^\bpl) Y^{kl}\cV_{kl}
+2u^{\opl i}v^{\bpl I}Y_{(i}{}^{l}\cC_{I)l}
~,~~~~~~~~
\label{de-deb+-2}
\eea
where, given any $A^{ij}$ with two SU(2) indices,
we use  $A^{\opl\bpl}=u_i^{\opl}v_I^{\bpl}A^{iI}$ as a contraction rule.
The following relations hold
\bsubeq
\bea
u^{\opl i}v^{\bpl I}{\bm L}_{iI}U^{(m,n)}= 
-\frac{m}{2}(u^\opl v^{\bpl}) U^{(m,n)} ~&,&
~~
u^{\opl i}v^{\bpl I}{\bm R}_{iI}U^{(m,n)}= \frac{n}{2} (u^{\opl}v^\bpl) U^{(m,n)} ~,~~
\label{usf111111}
\\
\cV^{\opl\bpl}U^{(m,n)}=
\frac{n-m}{2}(u^\opl v^{\bpl}) U^{(m,n)} 
~&,&~~
\cC^{\opl\bpl}U^{(m,n)}= -\frac{m+n}{2} (u^{\opl}v^\bpl) U^{(m,n)} ~.~~~~~~
\eea
\esubeq
Moreover, one finds that the combination 
$\big((u^\opl v^\bpl) A^{kl}\cV_{kl}-u^{\opl i}v^{\bpl I}A_{(i}{}^{l}\cC_{I)l}\big)$,
when acting on a bi-isotwistor superfield,
satisfy the simple relation
\bea
\big((u^\opl v^\bpl) A^{kl}\cV_{kl}-u^{\opl i}v^{\bpl I}A_{(i}{}^{l}\cC_{I)l}\big)U^{(m,n)}=
(m-n)A^{\opl\bpl}U^{(m,n)}
~.
\label{usf222222}
\eea
Using the last results and $\cm U^{(m,n)}=\frac{(m-n)}{2} U^{(m,n)}$, 
one easily obtains
\bea
\{\de_{+}^\opl,\de_{-}^\bpl\}U^{(m,n)}&=&
0~.
\eea
 
Now, we consider the anticommutator $\{\de_{+}^\opl,\deb_{-}^\bpl\}$
which can be easily seen to be
\bea
\{\de_{+}^\opl,\deb_{-}^\bpl\}&=&
4(u^\opl v^\bpl)\cT\cM
+4\cT\cV^{\opl\bpl}
-4\ri (u^\opl v^\bpl)\cS\cM
-4\ri\cS\cV^{\opl\bpl}
\non\\
&&
-4\ri\cT^{\opl\bpl}\cM
+2\ri (u^\opl v^\bpl)\cT^{kl}\cV_{kl}
-2\ri u^{\opl i}v^{\bpl I}\cT_{(i}{}^l\cC_{I)l}
\non\\
&&
+4\cS^{\opl\bpl}\cM
-2(u^\opl v^\bpl)\cS^{kl}\cV_{kl}
+2u^{\opl i}v^{\bpl I}\cS_{(i}{}^l\cC_{I)l}
~.
\eea
Therefore, by using (\ref{usf111111})--(\ref{usf222222}), 
one finds
\bea
\{\de_{+}^\opl,\deb_{-}^\bpl\}U^{(m,n)}=0~.
\eea

Along the same lines, it can be proved that
\bea
\{\deb_{+}^\opl,\deb_{-}^\bpl\}U^{(m,n)}&=&0~,~~~~~~
\{\deb_{+}^\opl,\de_{-}^\bpl\}U^{(m,n)}=0~.
\eea
This concludes the derivation of the equations (\ref{dedeU}).


\begin{small}

\end{small}

\end{document}